\documentclass[%
reprint,10pt,
superscriptaddress,
amsmath,amssymb,
aps,prl
]{revtex4-2}
\usepackage[margin=0.8in]{geometry}
\usepackage[utf8]{inputenc}
\usepackage[hidelinks]{hyperref}
\usepackage{graphicx}
\usepackage{amsmath,amssymb}
\usepackage{caption}
\usepackage{float}
\makeatletter
\let\newfloat\newfloat@ltx
\makeatother
\usepackage{algorithm}
\usepackage{algpseudocode}

\usepackage{wrapfig,lipsum}
\usepackage{float}

\usepackage{appendix}
\usepackage{titletoc}

\usepackage{hyperref}
\usepackage{xcolor}

\newif\ifcomment
\def\comment#1{%
    \ifcomment\relax\else #1\fi}

\begin{document}
\title{The Geometry of Jamming Algorithms\\ in the Random Lorentz Gas}
\author{Giampaolo Folena}
\affiliation{Dipartimento di Fisica, Sapienza Università di Roma, Piazzale Aldo Moro 5, Roma, 00185 Italy}
\email{giampaolofolena@gmail.com}
\author{Patrick Charbonneau}
\affiliation{Department of Chemistry, Duke University, Durham, North Carolina 27708}
\affiliation{Department of Physics, Duke University, Durham, North Carolina 27708}
\author{Peter K. Morse}
\affiliation{Department of Physics, Seton Hall University, South Orange, NJ 07079}
\author{Rafael Díaz Hernández Rojas}
\affiliation{Institute for Theoretical Physics, Georg-August-Universität Göttingen, Göttingen, Germany}
\author{Federico Ricci-Tersenghi}
\affiliation{Dipartimento di Fisica, Sapienza Università di Roma, INFN -- Sezione di Roma1, and CNR -- Nanotec, Piazzale Aldo Moro 5, Roma 00185 Italy}

\begin{abstract}\subsection*{Abstract:} 
Deterministic optimization algorithms unequivocally partition a complex energy landscape in inherent structures (ISs) and their respective basins of attraction. 
Can these basins be defined solely through geometric principles? This question is paramount to understanding hard sphere jamming, a key model of disordered matter.  We here address the issue by proposing a geometric class of gradient descent--like algorithms, which we use to study a system in the hard-sphere universality class, the random Lorentz gas. The statistics of the resulting ISs is found to be strictly inherited from those of Poisson--Voronoi tessellations. The landscape roughness is further found to give rise to a hierarchical organization of ISs, which various algorithms explore differently. In particular, greedy and reluctant schemes tend to favor ISs of markedly different densities. The resulting ISs nevertheless robustly exhibit a universal force distribution, thus confirming the geometric nature of the jamming universality class. Along the way, the physical origin of a dynamical Gardner transition is identified. %

\end{abstract}

\maketitle 

\subsection*{Introduction}
Jamming granular systems -- either for sand play or industrial transport~\cite{frenkel2010eggs} -- ubiquitously gives rise to disordered materials; so does supercooling many liquids.
Over a quarter of a century ago, this analogy led Liu and Nagel to  propose a unification of the two processes under a single conceptual umbrella~\cite{liu1998jamming}. Their \emph{jamming phase diagram} has since inspired fields as diverse as robotics~\cite{manti2016stiffening}, tissue mechanics~\cite{lawson2021jamming}, and neural networks~\cite{spigler2019jamming,dAscoli2020double}. It has also seeded a substantial research effort aiming to flesh out the original proposal itself~\cite{liu2010jamming,berthier2011theoretical,torquato2010jammed}. 

In this last respect, particularly significant strides have been made from the study of simple liquids in the limit of infinite spatial dimensions, $d\rightarrow\infty$~\cite{parisi2020theory}. 
This seemingly abstract construction has indeed rationalized jamming marginality 
and its associated isostaticity~\cite{wyart2012marginal,lerner2013low,muller2015marginal}, and made stunningly accurate predictions about the non-trivial scaling of the distribution of weak interparticle forces $P_F(f)\sim f^\theta$ and small interparticle gaps $P_H(h)\sim h^{-\gamma}   $ down to $d=2$~\cite{donev_pair_2005,skoge2006packing,
lerner2013low,charbonneau2015jamming,charbonneau2021finitesize,babu2022criticality,wang2022}. Despite jamming being an inherently out-of-equilibrium and hence a protocol dependent phenomenon, its criticality is seemingly universal. That robustness, however, remains largely unexplained. 

Another -- in some ways more salient -- theoretical challenge entails predicting the jamming density. For three-dimensional hard spheres, that quantity had long been understood to be algorithm invariant, with a volume fraction of about 64\%, thus cementing \emph{random close packing} (RCP) as a physically robust and universal concept. Over the last couple of decades, however, the confounding role played by various factors, such as the degree of crystallinity~\cite{torquato2000mrj} and the preparation scheme~\cite{
charbonneau2021finitesize,morse2024amorphous}, including details of the initial conditions~\cite{ozawa2012jamming,charbonneau2021finitesize}, have softened confidence in the robustness of RCP. At this point, even identifying a physically (let alone mathematically~\cite[p.~240--2]{chiu2013stochastic}) meaningful substitute remains an open challenge. In light of this, several alternatives have been offered as replacements for RCP: numerical extensions of the equilibrium liquid line~\cite{Kamien2007}, the endpoint of absorbing-state processes~\cite{Wilken2021}, the most random jammed configuration~\cite{torquato2000mrj}, and the densest isostatic packing~\cite{boltonlum2024}. While these approaches are useful in their own contexts, all introduce some degree of algorithmic arbitrariness such as explicit or implicit thermalization~\cite{morse2024amorphous}.

Some of us have recently proposed that ``the lowest jammed density achievable through bulk physical processes involving monotonic compression'', $\varphi_\mathrm{J0}$, might be a well defined quantity~\cite{morse2024amorphous}.
Put differently, to determine $\varphi_\mathrm{J0}$ one needs to identify an algorithm for jamming hard spheres akin to gradient descent (GD) for energy minimization in that it is both greedy and local. 

This idea is not new. In the mid-1980s, Stillinger and Weber systematically approximated the jammed inherent structures (IS) of hard spheres by using GD for systems with ever steeper interactions~\cite{stillinger1985}. 
More recently, Torquato and Jiao have formulated a linear programming scheme for hard spheres to reach jamming~\cite{torquato2010robustLP} -- later generalized as CALiPPSO~\cite{artiaco2020,artiaco2022} -- and Lerner et al. have formulated  an overdamped compression scheme for that same purpose~\cite{lerner2013low}. For various reasons, however, the effectiveness of these schemes at attaining $\varphi_\mathrm{J0}$ has not been systematically evaluated.  (See Ref.~\cite{morse2024amorphous} for an effort along this direction.) More problematically, a qualitative understanding of the similarities and differences between these schemes has yet to be teased out. Even in the simplifying limit of $d\rightarrow\infty$ (out-of-equilibrium) insights are in short supply~\cite{manacorda2022gradient}.


In order to shed a new light on both jamming robustness and density, we here consider the random Lorentz gas (RLG), a single-particle model that belongs to the hard sphere universality class in the limit $d\rightarrow\infty$~\cite{biroli2021interplay}. But what is jamming -- an ostensibly collective phenomenon -- in a single-particle system? 
Inspired by a recent geometrical study~\cite{Bonnet2024} and connections between the Voronoi network and jamming~\cite{morse2014}, we identify jammed ISs as inscribed spheres and propose a class of GD-like volume ascent (VA) algorithms for identifying them. 
This approach determines $\varphi_\mathrm{J0}$ for the RLG and illuminates the algorithm dependence of jamming more generally. Remarkably, we also find that $P_F(f)$ converges to the anomalous infinite-dimensional hard-sphere scaling for all algorithms, thus illuminating its geometrical origin. These advances not only pave the way for a deeper understanding of jamming but also hold the potential for improving high-dimensional optimization with hard constraints, which notably appear in the fields of computational geometry~\cite{de2000computational,preparata2012computational,toth2017handbook} and robust optimization~\cite{ben2009robust}.

\begin{figure}[!ht]
		\includegraphics[width=\columnwidth]{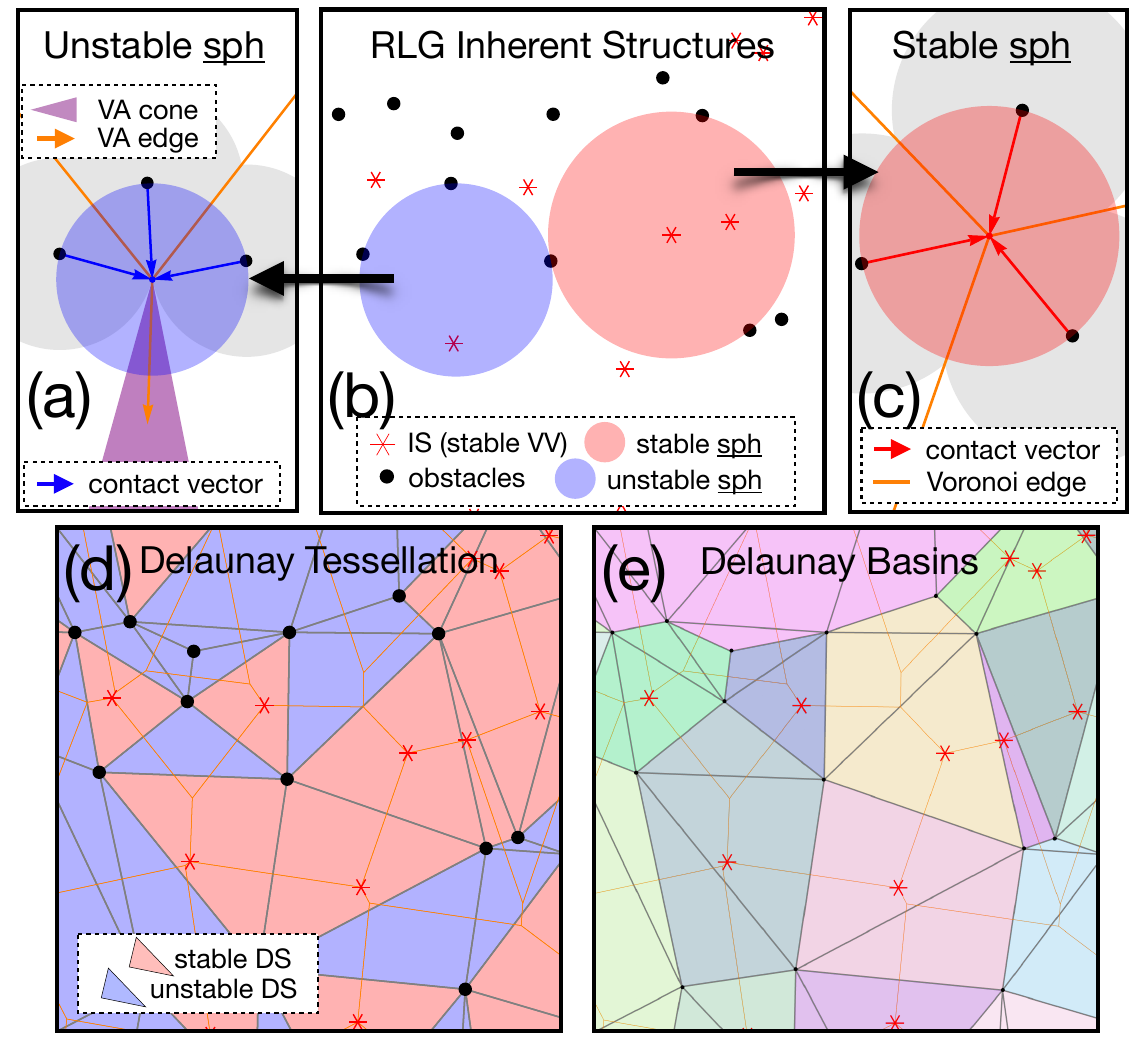}\\
		\caption{\textbf{(b):} Geometric (or entropic) landscape of the $d=2$ RLG with obstacles (black dots) and ISs (red stars). A sph is in contact with  $d+1$ obstacles. It is unstable if these contacts are co-hemispheric (light blue) and stable otherwise (light red).
        Contact vectors for the \textbf{(a)} unstable  and \textbf{(c)} stable  sph in (b), along with the VA cone of possible expansion directions (purple) for the former.
        \textbf{(d):} Delaunay tessellation of the sample in (b). An unstable DS does not contain its circumcenter (blue), while a stable DS does (red), thus identifying an IS (stars).
        \textbf{(e):} Delaunay basins from Eq.~\ref{eq:DelB} (different colors) of the sample in (b). These basins are generally composed of one stable DS and zero or more surrounding unstable DS. The DSs (black lines) and the Voronoi tessellation (orange lines) are provided as reference.}
  \label{fig:Fig1}
\end{figure}

\subsection*{The Entropic Landscape of the RLG}
Recall that the RLG consists of one spherical tracer evolving in the space unoccupied by hard (yet non-interacting) fixed spherical obstacles. These obstacles, which are distributed uniformly at random, form a Poisson process with number density (or intensity) $\rho=N/V$. One RLG convention assigns both tracer and obstacles the same sphere radius $r/2$; equivalently, one could consider a point tracer and obstacles of radius $r$, or point obstacles and a tracer of radius $r$. To ease visualization, we here mainly follow this last convention. 
To make densities unitless and of order one in all $d$, we further set $\rho=1$ and define the reduced volume fraction $\hat{\varphi}=\rho V_d r^d/d = V_d r^d/d$ with $V_d$ the volume of a $d$-dimensional unit sphere~\cite{biroli2021interplay}. 

To identify the ISs of the RLG, we follow the approach developed for random polytopes in Ref.~\cite{Bonnet2024}. At equilibrium, $r$ is constant; compression inflates $r$. At each tracer position $\mathbf{x}$ there exists a maximal sphere (\text{sph}) of radius $r_\mathrm{sph}$ that does not overlap with any obstacles. An IS is obtained when that radius $r_\mathrm{sph}$ can no longer locally grow by changing position $\mathbf{x}$, without creating overlaps with obstacles, hence determining $r_\mathrm{IS}$. The resulting tracer position, $\mathbf{x}_\mathrm{IS}$, is then equidistant from $d+1$ obstacles: fewer would not ensure mechanical stability, and more -- given the Poissonian nature of the obstacles -- would be highly improbable. Per the Maxwell criterion, the resulting IS is therefore isostatic.
From this analysis, an unambiguous determination of all IS for a given obstacle distribution follows (see Fig.~\ref{fig:Fig1}b). In geometrical terms, each point equidistant to $d+1$ obstacles is a  Voronoi vertex (VV) or, equivalently, a circumcenter of the dual Delaunay tessellation. As shown in Ref.~\cite{morse2023local}, a VV is \textit{stable} if it is contained within the respective Delaunay simplex (DS), and \textit{unstable} if not. Only the former are IS (see Fig.~\ref{fig:Fig1}(a-c)). 


\subsection*{Number of IS}
\begin{figure}[!ht]

		\includegraphics[width=\columnwidth]{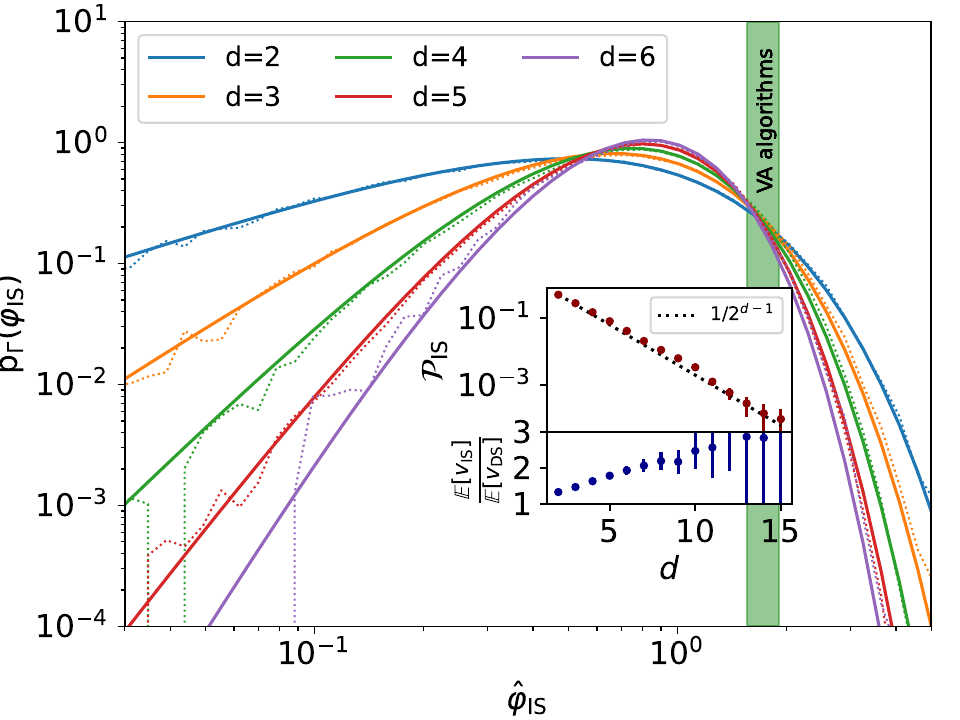}

		\caption{Probability distribution of ISs with packing fraction $\hat{\varphi}_\mathrm{IS}$ in $d=2\ldots6$. The expression in  Eq.~\ref{eq:cIS} (full lines) agrees with the numerical enumeration of stable ISs (dotted lines). Note that the distribution peaks at $\mathrm{mode}[\hat{\varphi}_\mathrm{IS}]=(d-1)/d$, and that 
        the expected jamming packing fraction is $\mathbb{E}[\hat{\varphi}_\mathrm{IS}]=1$ for all $d$. VA algorithms, by contrast, typically reach significantly larger values, as shown here for $d=2\dots 6$ (green band).  (inset) The proportion of stable DSs (points) empirically scales as $1/2^{d-1}$ (dotted line), while the ratio between the expected volume of stable DSs and the expected volume of all DSs grows sublinearly in $d$.}
          \label{fig:Fig2}
\end{figure}
Having established that each IS is the circumcenter of a stable DS, we now obtain a count of these features (per unit volume), $\mathcal{N}_\mathrm{IS}$, as a function of their packing fraction at jamming, $\hat{\varphi}_\mathrm{IS}$ (or, equivalently, $r_\mathrm{IS}$). This analysis therefore provides a measure of configurational entropy of ISs, $\Sigma_\mathrm{IS}=\ln \mathcal{N}_\mathrm{IS}$.

A result from stochastic geometry~\cite{edelsbrunner2017expected}, which holds for unbounded systems, gives that the density of DS -- both stable and unstable -- of a given $\hat{\varphi}$ per unit volume $\mathcal{N}_\mathrm{DS}(\hat{\varphi})$ is proportional to the Gamma probability distribution function $p_\mathrm{\Gamma}(\hat{\varphi})=d^d e^{-d\hat{\varphi}} \hat{\varphi}^{d-1}/\Gamma(d)$, i.e., 
\begin{equation}\label{eq:cDS}
    \mathcal{N}_\mathrm{DS}(\hat{\varphi})= N_\mathrm{DS}(d) p_\mathrm{\Gamma}(\hat{\varphi}),
\end{equation}
where $N_\mathrm{DS}(d)$ is the average number of DSs per unit volume\footnote{Low-$d$ results are known~\cite{edelsbrunner2017expected}: $N_\mathrm{DS}(d=2)=2$, $N_\mathrm{DS}(3)=6.76\ldots$, $N_\mathrm{DS}(4)=31.77\ldots$.}. 
Remarkably, given~\cite[Thm.~(10.4.4)]{schneider2008}, the geometry of each DS is statistically independent of its radial size -- or, equivalently,  the stability condition is statistically
independent from the radial size of the cell -- and hence 
\begin{equation}\label{eq:cIS}
    \frac{\mathcal{N}_\mathrm{IS}(\hat{\varphi})}{ N_\mathrm{IS}(d)}=\frac{\mathcal{N}_\mathrm{DS}(\hat{\varphi})}{ N_\mathrm{DS}(d)}= p_\mathrm{\Gamma}(\hat{\varphi})\ .
\end{equation}
The number of ISs per unit volume $N_\mathrm{IS}(d)=\mathcal{V}_\mathrm{IS}(d)N_\mathrm{DS}(d)$, however, is strongly reduced by a proportionality factor that can be decomposed as
$\mathcal{V}_\mathrm{IS}(d)=\mathcal{P}_\mathrm{IS}(d)\mathbb{E}[v_\mathrm{IS}]/\mathbb{E}[v_\mathrm{DS}]$, where $\mathcal{P}_\mathrm{IS}(d)$ is the fraction of stable DSs and $\mathbb{E}[v_\mathrm{IS}]/\mathbb{E}[v_\mathrm{DS}]$ is the ratio of the expected volume of stable DSs to the expected volume of all DSs.

Figure~\ref{fig:Fig2} shows that numerical results obtained using the standard qhull library~\cite{qhull} for Voronoi tessellation agree with ~\eqref{eq:cDS} and  ~\eqref{eq:cIS}, as expected. Because $\int\! d \hat{\varphi}\; \hat{\varphi} p_{\Gamma}(\hat{\varphi})=1$ for all $d$, we then have that sampling ISs uniformly at random gives $\mathbb{E}_\mathrm{IS}[\hat{\varphi}]=1$ for all $d$.  Numerics further reveal that $\mathcal{P}_\mathrm{IS}(d)$ is exponentially suppressed in $d$, approximately scaling as $1/2^{d-1}$,\footnote{This ratio is akin to Wendel's theorem, which predicts that $1/2^{d}$ of simplexes are stable when sampling $d+1$ vertices on a $d$-dimensional sphere. The two cases, however, differ because random Delaunay simplexes tend to be more stable than random simplexes.} while $\mathbb{E}[v_\mathrm{IS}]/\mathbb{E}[v_\mathrm{DS}]$ grows slower than exponential with $d$ (Fig.~\ref{fig:Fig2} inset). In other words, in high $d$, unstable DSs exponentially dominate both in number and in volume over stable DSs. 

\subsection*{VA Algorithms and Geometry}
Despite the scarcity of stable DSs, some algorithms can nevertheless reach them in polynomial time. To this effect, we here specifically define volume ascent (VA) schemes as the class of \emph{local} algorithms which monotonically inflate $r_\mathrm{sph}$. Note that the VA problem can be formulated more generally within the mathematical framework of non-smooth optimization \cite{clarke1990optimization,gaudioso2020essentials}. Specifically, $r_\mathrm{sph}$ is a continuous but non-differentiable function of $\mathbf{x}$, thus defining a rough landscape with many kinks, for which standard optimization techniques are ineffectual. Our analysis of stable DSs, therefore, provides insight into the dynamics of these algorithms more generally.

For VA schemes, first, each unstable Delaunay simplex is associated with a Voronoi vertex that presents a cone of possible VA directions (VA cone), within which the edges of the Voronoi tessellation identify a subset of specific VA pathways (VA edges). While in $d=2$ an unstable Voronoi vertex contains only one such VA edge (see Fig.~\ref{fig:Fig1}a), for $d>2$ that number varies between $1$ and $d-1$.

Second, because the volume fraction of stable DSs is exponentially suppressed with increasing $d$,
the system volume is almost completely filled with unstable DSs in the large $d$ limit. Any VA algorithm must therefore flow between many unstable VVs before reaching an IS. Because VVs can be hierarchically connected through VA edges, selecting a specific VA algorithm is equivalent to selecting one such compression pathway (either along VA edges or in between them).

\begin{figure}[!ht]

	\includegraphics[width=0.98\columnwidth]{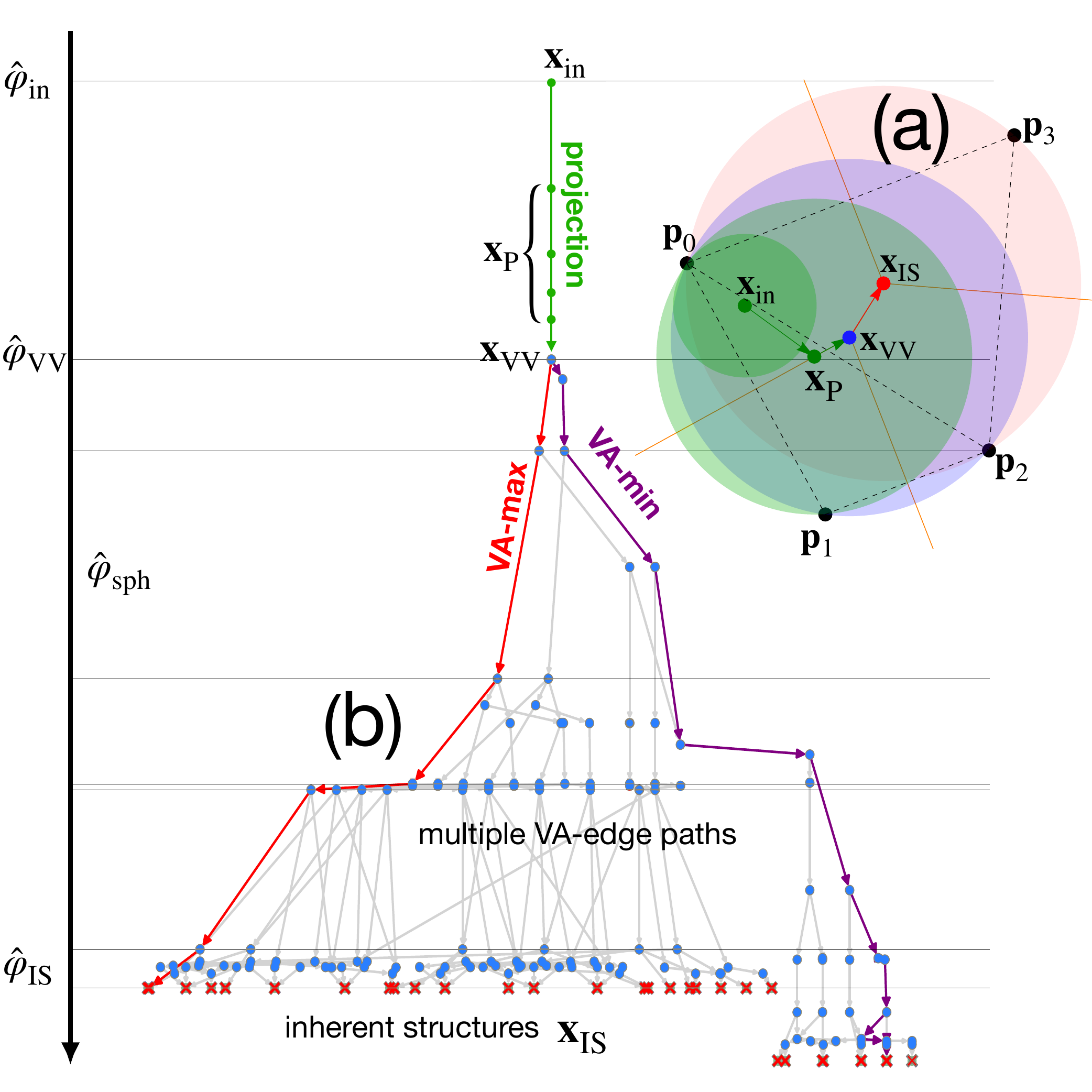}
        \caption{
\textbf{(a):} Schematic of the VA-max algorithm for the $d=2$ RLG. The tracer starts at $\mathbf{x}_\mathrm{in}$, moves radially from $\mathbf{p}_0$ until it reaches the Voronoi line at $\mathbf{x}_\mathrm{P}$, and then follows it until reaching the first VV at $\mathbf{x}_\mathrm{VV}$. It subsequently follows the VA edge until $\mathbf{x}_\mathrm{IS}$, which corresponds to a vertex of the Voronoi tessellation. \textbf{(b):} In generic $d$, the VA cone at each VV can present up to $d-1$ VA edges, thus giving rise to the multi-path structure of the VA-edge algorithm. After the initial $d$-step projection up to $\hat\varphi_\mathrm{VV}$ (green), multifurcation along various VA edges results in different IS (red crosses). Path coalescence can also arise, but is rare. The VA-max (red) and the VA-min (purple) paths are the greediest and the most reluctant VA-edge algorithms, respectively. For this $d=5$ example, the abscissa is chosen so as to minimize path crossings.}
  \label{fig:Fig3}
\end{figure}

Among all possible VA algorithms, we first consider the \textit{greediest} one, VA-max.
The tracer displacement, $\mathbf{dx}$, is then  chosen to maximize the growth of the sph radius, $r_\mathrm{sph}$, at each position $\mathbf{x}$,
\begin{equation}\label{eq:Rmax}
\mathbf{dx} = \underset{\mathbf{d\tilde{x}}}{\text{argmax}}\; r_\mathrm{sph}(\mathbf{x} + \mathbf{d\tilde{x}}),
\end{equation}
thus making it a direct analog of GD for energy minimization.
(For a broader definition of steepest descent in non-smooth landscapes, see Ref.~\cite{gaudioso2020essentials}.) Figure~\ref{fig:Fig3}(a) illustrates the process for $d=2$.
Starting from a point tracer originally at $\mathbf{x}=\mathbf{x}_\mathrm{in}=\Vec{0}$, the maximal radius available $r_\mathrm{sph}$ is equal to the distance to the closest obstacle $|\mathbf{p}_0|$. Following Eq.~\ref{eq:Rmax}, the tracer then moves radially from the closest obstacle ($\mathbf{dx}\propto -\mathbf{p}_0$) while its radius grows as $r_\mathrm{sph}= |\mathbf{x}-\mathbf{p}_0|$, until the tracer kisses a second obstacle at $\mathbf{p}_1$. Its center is then equidistant from $\mathbf{p}_0$ and $\mathbf{p}_1$. The subsequent dynamics follows the Voronoi hyperplane defined by points equidistant from both $\mathbf{p}_0$ and $\mathbf{p}_1$, until the sph reaches a third obstacle at $\mathbf{p}_2$. In $d=2$, the tracer is then equidistant from $d+1=3$ obstacles, and its center is on a Voronoi vertex (VV), by definition. In general $d$, the center of the sph reaches a VV by $d$ Gram--Schmidt projections of the initial growth direction $-\mathbf{p}_0$ onto $\text{span}\{\mathbf{p}_1,\dots,\mathbf{p}_{d}\}$, thus defining $\hat{\varphi}_\mathrm{VV}$.\footnote{This process is equivalent to projecting the initial point onto one of the vertexes of the Voronoi polytope defined by points that are closer to $\mathbf{x}_1$ than to any other obstacle.}

If this VV is stable, then an IS has been reached and the dynamics stops. Otherwise, the trajectory flows in the greediest direction inside the VA cone, which necessarily lies along one of the VA edges (see Fig.~\ref{fig:Fig1}(a)). For $d=2$, because the VA cone contains only one such edge, VA-max follows it up the next Voronoi vertex. In $d>2$ at each VV the greediest VA edge is followed until another VV is found, and so on (SI Appendix). 
Figure~\ref{fig:Fig3}(b) illustrates the VA-max algorithm among the graph of VA edges connecting VVs. All other paths are \textit{reluctant} VA-edge schemes. Among those, we also define the most reluctant VA-min, which selects the direction of least growth at each VV. (VA-min, however, is not the most reluctant algorithm of the overall VA class, for which an inf does not exist.) As expected for systems with complex landscapes \cite{parisi2003statistical}, more reluctant algorithms typically achieve larger packing fractions $\hat{\varphi}_\mathrm{IS}$, over a larger number of steps than VA-max (SI Appendix). 
The graph of all possible VA-edge paths from the initial projection $\mathbf{x}_\mathrm{VV}$ to all final ISs also presents two key features of rough landscapes: \textit{multifurcation} and \textit{coalescence}. The former is directly connected to the presence of multiple VA edges at each VV and its probability grows with $d$;  the latter is the convergence of two trajectories that have previously bifurcated and its probability vanishes as $d$ increases.

 \begin{figure}[!ht]

		\includegraphics[width=0.95\columnwidth]{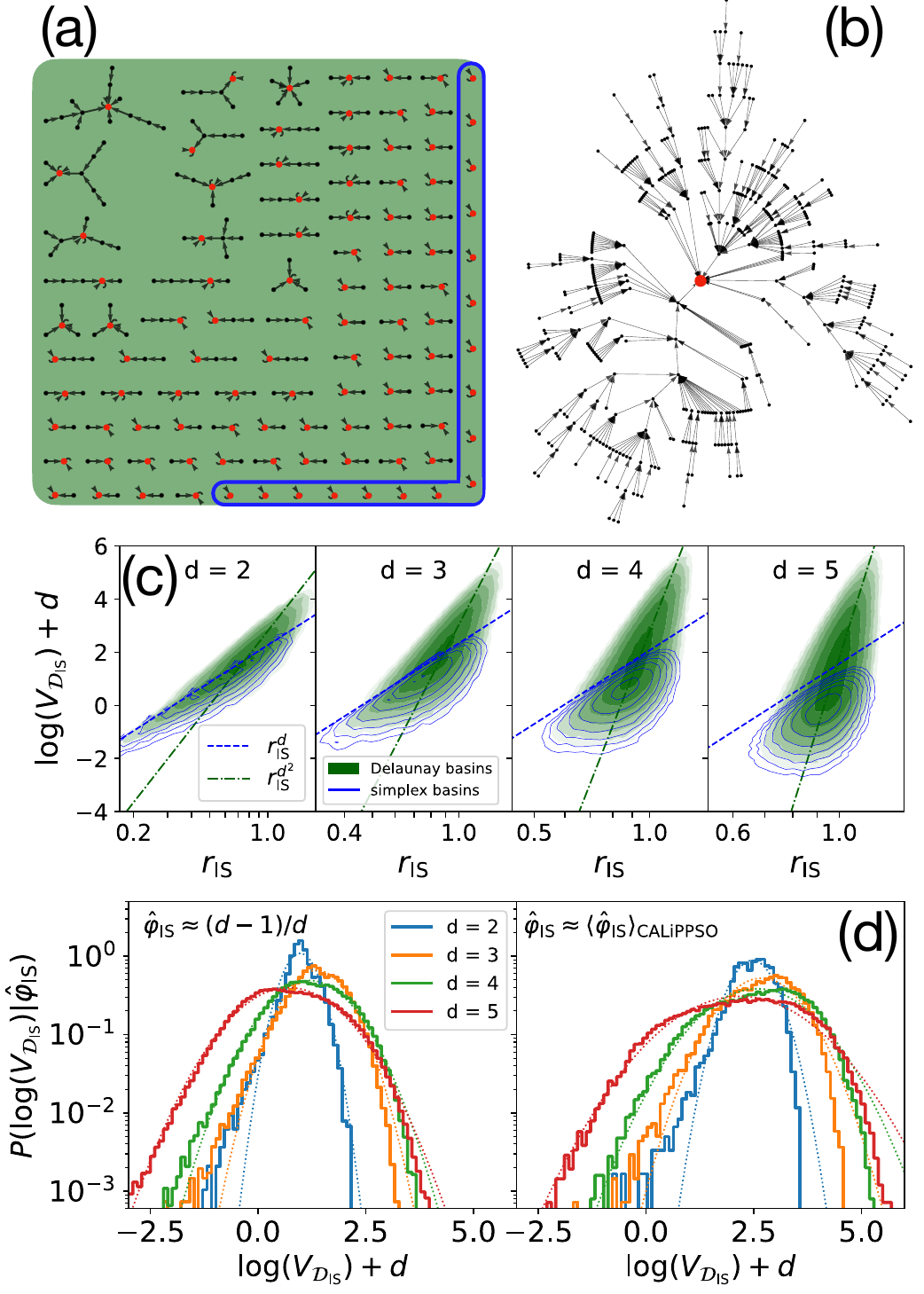}
        \caption{\textbf{(a):} Tree graphs of the Delaunay basins in Fig.~\ref{fig:Fig1}e as described in the text. The fraction of basins with a single DS (circled in blue) decreases exponentially as $d$ increases. 
        \textbf{(b):} Tree graph of a large Delaunay basin in  $d=4$ illustrating the growing fractal-like nature of these basins with $d$. \textbf{(c):} Joint distribution of the logarithm of the Delaunay basin volume $\log(V_{\mathcal{D}_\mathrm{IS}})$ and the circumradius $r_\mathrm{IS}$ in $d=2\ldots5$ from direct Delaunay tessellation (SI Appendix). 
        The volume of basins composed of one simplex is upper bounded by the volume of regular simplexes (blue dashed line). 
        The non-compact dependence of the volume of Delaunay basins (green cloud) trends consistently with $\propto r_\mathrm{IS}^{d^2}$ (green dashed-dotted line), hinting at their fractal-like nature. Both blue and green clouds are pdfs with contour lines at $2^{k}$ with $k=-5\ldots1$. \textbf{(d):} Probability distribution of the logarithm of the Delaunay basin volume, conditioned on the packing fraction: the mode of the distribution, $\hat{\varphi}_\mathrm{IS} = (d-1)/d$, and a typical value obtained from the dynamics, $\hat{\varphi}_\mathrm{IS} = \langle \hat{\varphi}_\mathrm{IS} \rangle_\mathrm{CALiPPSO}$. Log-normal fits  (dotted lines) are provided as reference.}
  \label{fig:Fig4}
\end{figure}

\subsection*{Fractal Basins}\label{par:frac}
The basin of attraction $\mathcal{B}^\mathrm{alg}_\mathrm{IS}$ for a given (deterministic) algorithm is the ensemble of all initial configurations $\mathbf{x}_\mathrm{in}$ that reach a same IS. Its volume is proportional to the algorithmic probability to sample that IS. For VA-max in $d=2$, a geometrical analysis of these basins is straightforward. Because every trajectory starting within a given DS flows towards the VV at its circumcenter, the flow of trajectories clusters all DSs that flow to the same IS. The basin of attraction of each IS is therefore a pure \emph{Delaunay basin}
\begin{equation}\label{eq:DelB}
\mathcal{D}_\mathrm{IS} = \{a \;|\; \text{circ}(a) = \mathbf{x}_\mathrm{IS} \;\text{or}\; \exists b \in \mathcal{D}_\mathrm{IS} \; \text{s.t.} \; \text{circ}(a)  \subset b \} \ ,
\end{equation} 
where $a,b$ are DS and $\text{circ}(a)$ is the circumcenter of $a$.
Put differently, the basins of attraction can be decomposed as one stable DS surrounded by unstable DSs. As can be seen in Fig.~\ref{fig:Fig1}e, in $d=2$ the resulting basins are heterogeneous in shape and, in contrast to the cells in Voronoi or Delaunay tessellations, not necessarily convex.
From the definition of a Delaunay basin, a simple greedy algorithm naturally follows: at each step, the tracer moves to the circumcenter defined by the vertices of the Delaunay simplex (DS) that currently contains it. In other words, it displaces the tracer to a nearby Voronoi vertex that most increases its radius (SI Appendix).  
Remarkably, in the RLG, this procedure is equivalent to the CALiPPSO linear optimization algorithm~\cite{artiaco2020,artiaco2022}. As a result, basins of attraction for this scheme exactly correspond to Delaunay basins in all $d$. 
For VA-max in $d>2$, although a similar decomposition does not exactly describe  basins due to the initial simplex having more than one unstable direction, numerical results suggest it is nevertheless a very good approximation, especially for large $d$ (SI Appendix). 

The basin structure of any greedy algorithm naturally forms a tree-like graph. 
The root vertex is then the stable DS, each node corresponds to a simplex, and each edge represents the algorithmic step from one simplex to another. 
Figures~\ref{fig:Fig4}(a)-(b) depict the tree-like organization of DSs within a Delaunay basin, derived using the Eq.~\ref{eq:DelB} definition, after numerically tessellating the space into DSs. Given that the number of simplexes within a typical basin scales exponentially with $d$ (due to the above-mentioned scarcity of stable DSs), we expect that, even in moderately high dimensions, basins exhibit intricate graph structures with non-trivial fractal dimensions (SI Appendix).

To further characterize this structure, we consider the joint distribution of basin volume $V_{\mathcal{D}_\mathrm{IS}}$ and circumcenter radius $r_\mathrm{IS}$ (or equivalently, packing fraction) in Fig.~\ref{fig:Fig4}(c). In all $d$, the basin volume growth is consistent with $\propto r^{d^2}$, much faster than for compact objects, for which $V\propto r^d$. (Recall that the RLG is a single-particle system, and hence its phase space is $d$-dimensional.) 
Figure~\ref{fig:Fig4}(d) presents the log-distribution of volumes conditioned on packing fraction for $d=2,3,4,5$. Two packing fractions are considered: (i) the mode of the uniform sampling distribution, 
and (ii) the typical value reached dynamically, $\langle \hat{\varphi}_\mathrm{IS} \rangle_\mathrm{CALiPPSO}$. In both cases, the distributions flatten as dimension increases. In all dimensions, the extension of the Edwards hypothesis to single-particle systems --namely, the assumption that all configurations at fixed $\hat{\varphi}_\mathrm{IS}$ are sampled uniformly at jamming~\cite{Edwards1989,Baule2018} -- appears to be violated. For the first packing fraction, the distribution is approximately log-normal, consistent with observations in multi-particle systems~\cite{Xu2011}. However, the distribution conditioned on the second packing fraction is significantly broader, indicating that the basins sampled by VA compression are highly heterogeneous in size. Interestingly, this behavior contrasts with what was reported for systems of polydisperse disks~\cite{Martiniani2017,Martiniani_Casiulis_2023}, for which the distribution narrows instead, consistent with the Edwards hypothesis. This discrepancy may point to a more fundamental distinction between single- and multi-particle systems. Perhaps the heterogeneity of single-particle basins is effectively smoothed out in multi-particle systems due to the presence of collective degrees of freedom that can be optimized. As a result, the basins of the latter might tessellate configuration space more efficiently than the former, thus suggesting a collective origin of the Edwards hypothesis. The precise origin of this difference remains, for now, an open question.

The basin volume distribution further provides insight into algorithmic outcomes. The basin volume of an algorithm indeed reweighs the contribution of an IS relative to the uniform distribution, 
\begin{equation}
    \langle \hat{\varphi}_\mathrm{IS}\rangle_\mathrm{alg}=\sum_{\mathrm{IS}} \frac{V_{\mathcal{B}^\mathrm{alg}_\mathrm{IS}}}{V_\mathrm{tot}} \hat{\varphi}_\mathrm{IS}
    \label{eq:reweigh}
\end{equation}
Because the uniform measure ensures $\mathbb{E}_\mathrm{IS}[\hat{\varphi}] = 1$, having $\langle \hat{\varphi}_\mathrm{IS}\rangle_\mathrm{alg}>1$ indicates a non-uniform weighting across basins. 

\begin{figure}[!ht]
		\includegraphics[width=0.99\columnwidth]{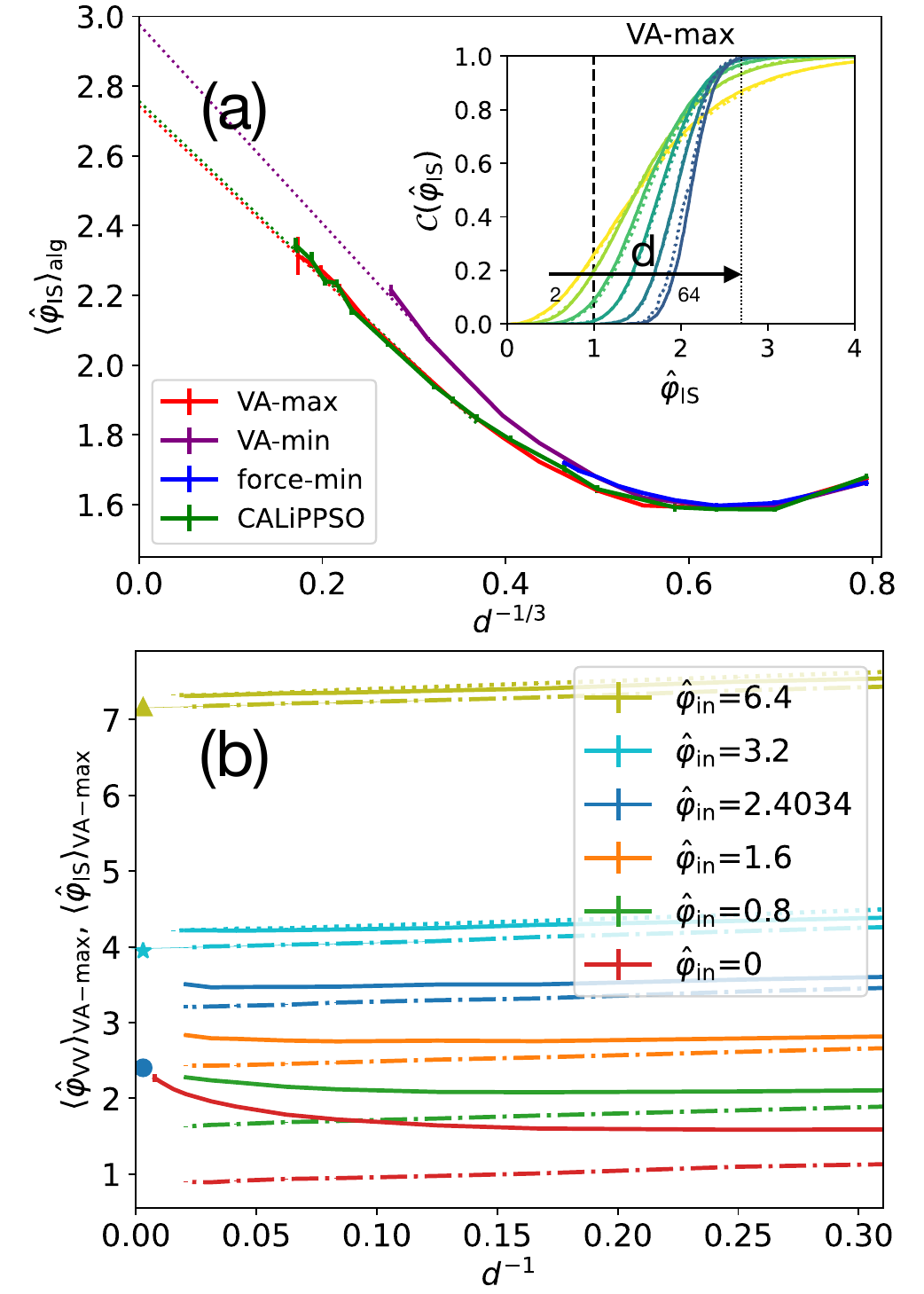}
		\caption{\textbf{(a):} Dimensional dependence of the jamming density reached from $\hat{\varphi}_\mathrm{in}=0$, $\langle\hat{\varphi}_\mathrm{IS}\rangle_\mathrm{alg}$, for VA-max, VA-min, force-min, and CALiPPSO. The empirical scaling  $d^{-1/3}$ gives a (nearly) linear scaling in large $d$. Both VA-max, the greediest local algorithm, and CALiPPSO, its non-local equivalent, achieve similar densities. Their extrapolation (thin dotted lines) suggests that $\hat{\varphi}_\mathrm{J0}=\langle\hat{\varphi}_\mathrm{IS}\rangle_\mathrm{VA\!-\!max}=2.73(2)$ in the limit $d\rightarrow\infty$. For all $d>2$, VA-min and force-min give significantly denser results, with $\langle\hat{\varphi}_\mathrm{IS}\rangle_\mathrm{VA\!-\!min}=2.94(9)$ asymptotically. \textbf{inset:} The cumulative distribution of $\hat{\varphi}_\mathrm{IS}$ for VA-max (full lines) in $d=2\ldots64$ markedly sharpens as $d$ increases. CALiPPSO results (dotted lines) are almost indistinguishable on this scale.
        The expected value for the uniform measure over ISs (dashed vertical line) as well as $\hat{\varphi}_\mathrm{J0}$ in the limit $d\rightarrow\infty$ (dotted vertical line) are given as reference.
         \textbf{(b):} Dimensional dependence of $\langle\hat{\varphi}_\mathrm{IS}\rangle_\mathrm{VA-max}$ (full lines) and of the finite-$d$ estimate of $\langle\hat{\varphi}_\mathrm{VV}\rangle_\mathrm{VA-max}$ (dashed-dotted lines) reached from various $\hat{\varphi}_\mathrm{in}$. 
         (The red curve is the same as in (a).) 
         For $\hat{\varphi}_\mathrm{in}\gg\hat{\varphi}_\mathrm{d}$, $\hat\varphi_\mathrm{G}$ (symbols) agrees with $\langle\hat{\varphi}_\mathrm{VV}\rangle_\mathrm{VA-max}$ obtained by a linear extrapolation  in $1/d$ (thin line), but not so for $\hat{\varphi}_\mathrm{in}\gtrsim\hat{\varphi}_\mathrm{d}$. Dotted lines indicate the VA-min curve for $\hat{\varphi}_\mathrm{in} = 3.2$ and $6.4$, closely matching the VA-max curve at these densities. 
         }\label{fig:Fig5}
\end{figure}

\subsection*{Jamming Density Results}
As expected for systems with complex landscapes (e.g. \cite{Folena2020,Folena_2021,Folena2023} for mean-field systems and \cite{Sastry1998,Nishikawa2022, Suryadevara2024} for multi-particle systems), different optimization algorithms reach ISs at different ``depths'' in the landscape. Although the maximal radius $r_\mathrm{IS}$ achieved when starting within the basin of attraction of a particular IS is geometrically fixed, the probability of ending in that  basin is algorithm dependent, as in Eq.~\ref{eq:reweigh}. We here consider the performance of three different VA algorithms, (i) VA-max,  (ii) VA-min, and (iii) force-min, as well as the VA-like scheme, (iv) CALiPPSO.
Recall that Lerner et al.'s force-min\cite{lerner2013simulations} 
is a VA reluctant algorithm that does not follow VV edges, while  CALiPPSO is a greedy but non-local scheme (SI Appendix). 
Figure~\ref{fig:Fig5}(a) compares  $\langle\hat{\varphi}_\mathrm{IS}\rangle_\mathrm{alg}$ obtained by averaging over both trajectories and realizations of disorder, starting from an initial density $\hat{\varphi}_\mathrm{in}=0$. In all cases, the results are markedly larger than the uniform sampling bound, $\mathbb{E}[\hat\varphi_\mathrm{IS}]=1$, mentioned above. As expected, VA-max gives the lowest jamming densities in all $d$, thus identifying $\hat{\varphi}_\mathrm{J0}=\langle\hat{\varphi}_\mathrm{IS}\rangle_\mathrm{VA-max}$ for $\hat{\varphi}_\mathrm{in}=0$; CALiPPSO gives nearly indistinguishable results. 
Remarkably, VA-min and force-min produce nearly identical results (albeit over a more limited $d$ range), just as VA-max and CALiPPSO do. The gap between these two pairs of algorithms, however, grows with $d$, thus underscoring the growing complexity of the optimization landscape.

In order to relate these finite-$d$ findings with the exact DMFT results for the limit $d\rightarrow\infty$ -- once they become available -- a dimensional extrapolation is needed. A simple $1/d$ scaling, however, does not capture the trend of even the highest $d$ results achieved, in marked contrast from what is observed for equilibrium observables in that same model~\cite{biroli2022local}. 
Recent results for other systems with complex landscapes suggest that altogether different finite-size scaling form, $d^{-a}$ with $a<1$, is to be expected~\cite{Folena_2021,erba2024}, but limited theoretical guidance is available for choosing $a$. 
Unfortunately, that choice substantially impacts the extrapolation outcome, with the systematic error far exceeding the statistical one. For instance, in the limit $d\rightarrow\infty$, $a=1/2$ gives $\hat\varphi_\mathrm{J0}=2.49(2)$ at one extreme, and $a=1/4$ gives $\hat\varphi_\mathrm{J0}=3.09(2)$ at the other (fits are for $d>30$). This density range is nevertheless largely consistent with earlier estimates obtained by GD on a softened RLG~\cite{manacorda2022gradient} and by force-min on the many-body problem~\cite{charbonneau2023jamming} (after appropriate rescaling~\cite{manacorda2022gradient}), especially given that these estimates were obtained from narrower $d$ ranges and extrapolated with $a=1$.

\subsection*{Gardner Transition}
Trajectories can also be started from finite initial (equilibrium) densities, i.e., $\hat{\varphi}_\mathrm{in}>0$. Algorithms are then expected to reach IS that lie deeper in the landscape, as shown for VA-max in Fig.~\ref{fig:Fig5}(b). Recall that in the limit $d\to\infty$, for $\hat{\varphi}_\mathrm{in}<\hat{\varphi}_\mathrm{d}=2.4034\ldots$ the equilibrium dynamics of the RLG is ergodic, while for $\hat{\varphi}_\mathrm{in}>\hat{\varphi}_\mathrm{d}$ it is localized to a cage. In the latter regime, a slow (adiabatic) compression as the tracer remains inside its cage makes \textit{state following} calculations possible (see \cite[Ch.~6]{parisi2020theory} for details). For each $\hat{\varphi}_\mathrm{in}$, there then exists a corresponding Gardner packing fraction $\hat{\varphi}_\mathrm{G}(\hat{\varphi}_\mathrm{in})$, at which the cage fractures into a full hierarchy of subcages, a transition known as full replica symmetry breaking (fullRSB). Simulations further hint that a Gardner-like transition might be generally observable in a fast (non-adiabatic) compression, but theoretical guidance is lacking~\cite{charbonneau2021finitesize}. 

The simplicity of the RLG landscape geometry makes a broader consideration of Gardner physics possible.
Recall that the initial VA projection collapses trajectories onto the first VV. In other words, a whole compact volume is reduced to a point. After this projection, the system explores the intricacies of the Voronoi edges that underlie landscape roughness (see Fig.~\ref{fig:Fig4}). The end of the projection phase at volume fraction $\hat{\varphi}_\mathrm{VV}$ therefore signals a dynamical transition analogous to the (adiabatic) Gardner transition in the limit $d\rightarrow\infty$, where the landscape becomes truly rough. Unlike the adiabatic transition, however, this dynamical transition can be observed for \emph{all} $\hat{\varphi}_\mathrm{in}$. 

Figure~\ref{fig:Fig5}(b) reports the dimensional evolution of $\hat{\varphi}_\mathrm{dG}=\langle \hat{\varphi}_\mathrm{VV}\rangle$, which in the limit $d\to\infty$ defines a dynamical Gardner transition. Unlike the jamming transition, $\hat{\varphi}_\mathrm{dG}$ clearly scales as $1/d$, consistent with its location being controlled by compact rather than rough landscape features. Its extrapolation to the limit $d\rightarrow\infty$ is therefore numerically robust. Interestingly, near $\hat{\varphi}_\mathrm{in}\gtrsim\hat{\varphi}_\mathrm{d}$ the adiabatic and the dynamical Gardner transitions are clearly distinct, with $\hat{\varphi}_\mathrm{dG}>\hat{\varphi}_\mathrm{G}$. For $\hat{\varphi}_\mathrm{in}\gg\hat{\varphi}_\mathrm{d}$, however, the two nearly coincide. In this regime, 
the tracer is caged within a nearly convex polytope, for which jamming is algorithm-independent. All algorithms -- be they greedy or adiabatic -- systematically converge to the center of the sphere inscribed in that polytope~\cite{Bonnet2024}. 
In this limit, the Gardner dynamical transition can then be defined as the point at which the shrinking cage loses its first facet (i.e. at the first Voronoi vertex). At this point, compression begins to experience bifurcations, thus mirroring the $d\rightarrow\infty$ Gardner transition, at which point the replica symmetric (i.e., convex) solution becomes unstable and gives way to a multiplicity of solutions.
For $\hat{\varphi}_\mathrm{in}<\hat{\varphi}_\mathrm{d}$, no adiabatic transition exists, but the dynamical one smoothly continues across. The density gap between the jamming and the dynamical Gardner transitions further grows as $\hat{\varphi}_\mathrm{in}$ decreases and persists even at $\hat{\varphi}_\mathrm{in}=0$. Finite-$d$ echoes of this physics should therefore be discernible along the compression trajectory of even the simplest of jamming systems, at least for $d>2$.

\begin{figure}[!ht]
    \includegraphics[width=0.98\columnwidth]{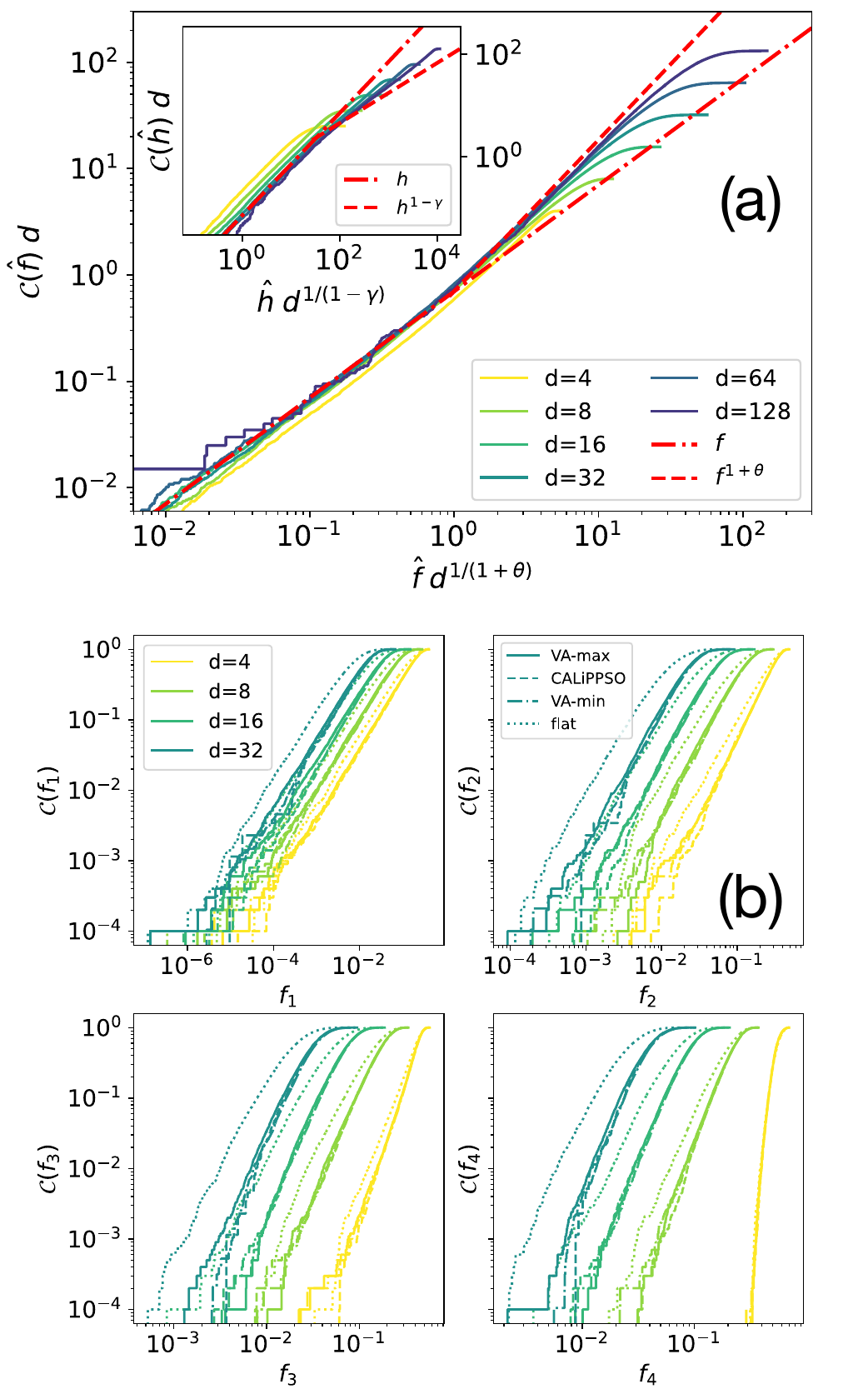}
		\caption{
        \textbf{(a):} Scaled cumulative force distribution for the rescaled force $\hat{f}=f/\langle f \rangle$ for VA-max ISs reached from $\varphi_\mathrm{in} = 0$ . 
        The rescaling results in a very good collapse and a clear crossover from the linear regime (dash-dotted line) to the anomalous power-law regime (dashed line), $f^{1+\theta\dots}$ with $\theta=0.4231\dots$.  
        (inset) The cumulative gap distribution does not similarly present an anomalous regime (with $\gamma=0.4127\dots$), consistent with the expected larger finite-$d$ correction (see text).
        \textbf{(b)}: Cumulative distribution of the smallest $f_k$ contact force for $k=1\ldots4$ for each IS in $d=4,8,16,32$. For each force type and each $d$, four different algorithms are considered: VA-max, CALiPPSO, VA-min, and uniform (flat) sampling. The resulting force distributions are identical for the first three, but differs for the uniform sampling. This supports the geometric universality observed in jamming physics.}
  \label{fig:Fig6}
\end{figure}

\subsection*{Jamming Universality}
Independent of the particular compression algorithm, isostatic jammed configurations have been reported to exhibit robustly universal properties. For many-body systems quantitative theoretical predictions obtained in the limit $d\to\infty$ \cite{parisi2020theory} have indeed been found to persist down to $d=2$  \cite{
charbonneau2015jamming,charbonneau2021finitesize}. Given that these critical scalings only emerge in the thermodynamic limit, this collective effect would not be expected to hold as-is for the single-particle RLG. The distribution of gaps between non-touching obstacles, for instance, remains far from the $d\rightarrow\infty$  scaling with $\gamma = 0.4127\dots$ even at the highest $d$ reached (see Fig.~\ref{fig:Fig6}(a):inset).  The small force distribution, however, does exhibit a power-law regime compatible with the predicted exponent, $\theta = 0.4231\dots$, even in fairly low $d$. This is confirmed  by the finite-size scaling analysis of Fig.~\ref{fig:Fig6}(a) for VA-max, using a squared-force normalization at each IS (see SI Appendix for details). 
The difference in $d$ scaling between the two observables is consistent with their finite-size dependence in many-body systems~\cite{charbonneau2021finitesize}. While the anomalous force distribution can be observed in fairly small systems, hints of the gap distribution require at least 200 particles.

More striking is that the geometry of isostatic contact vectors is robust even for small $d$. Different algorithms achieve IS with similarly distributed obstacles — an observation also made in multi-particle systems \cite{Wilken2021}. As shown in Fig.~\ref{fig:Fig6}(b), for instance, the cumulative distribution of the $k$-smallest force, $f_k$,  for $k=1\ldots4$ changes with $d$ but not with algorithm. VA-max, CALiPPSO, VA-min all give nearly indistinguishable results. Only the uniform sampling yields a distinct result. This difference can be interpreted -- by analogy with potential energy landscape analysis -- through the concepts of marginal and gapped IS. We conjecture that all sufficiently greedy (and sufficiently local) dynamics converge to marginal IS, and that all such marginal IS belong to a common jamming universality class, but a geometric definition of marginal IS in non-smooth energy landscapes is still lacking.
We conclude that the structural universality of jamming is present in small $d$, encoded by the landscape and independent of the specific dynamics. The algorithm merely sets the overall scale. A purely geometric analysis of marginal IS should therefore be able to extract the jamming critical exponents and hence fully explain their universality.
 
\subsection*{Conclusions}
In this work, we have studied the landscape geometry and the volume ascent class of optimization algorithms for a paradigmatic model of real-space jamming, the random Lorentz gas.  By analytically studying the complexity of its inherent structures, we have shown that in large dimensions, (phase) space is almost completely filled with volumes that are unstable under compression and we have identified the geometric origin of the ensuing landscape roughness. The basins of attraction therefore exhibit a growing hierarchical  and fractal structure as $d$ increases, a complexity reminiscent of what is reported for the many-body case~\cite{ashwin2012}. The greedy VA-max algorithm was further argued to be an optimal choice for computing $\varphi_\mathrm{J0}$. 

Through the landscape analysis, we have identified \emph{en passant}, a dynamical analogue of a Gardner transition that should be experimentally accessible, and shown that static analytical predictions for $d\rightarrow\infty$ agree with this definition for VA algorithms. We have also found that the structure of jammed configurations is unaffected by the choice of jamming algorithm and converges to mean-field predictions in the limit $d\to\infty$, indicating a geometric origin for jamming universality.

Several research directions stem from these results. First, a generalization of VA-max to multi-particle systems should be possible. It would then be interesting to consider whether these systems exhibit similar landscape properties as their single-particle counterpart. Given the lower computational complexity of the algorithm compared to previous proposals, a thermodynamic estimate of $\varphi_\mathrm{J0}$ -- a more physically and mathematically robust quantity than RCP  -- should then also be within reach. 
Second, solving the DMFT equations associated with the VA-max algorithm enables a direct investigation of the dynamics in the infinite-$d$ limit and a more accurate extrapolation of the asymptotic packing fraction. This contrasts with previous methods \cite{manacorda2022gradient}, which required first extrapolating to infinite time at finite softness and then taking the zero-softness limit—an indirect and potentially less reliable procedure. An exact solution of the scheme in the limit $d\to\infty$ should therefore be within reach. Third, the algorithmic robustness of jamming could be used to derive scaling laws from a purely geometrical approach.

In addition, the present work sheds new light on the broader class of real-space optimization problems in complex landscapes. ``Computing largest empty circles with location constraints'' \cite{toussaint1983computing}, for instance, is a fundamental problem in robust optimization \cite{ben2009robust,bertsimas2010robust,bertsimas2011theory}. Although heuristic algorithms can approximate its optimal solution in any dimension \cite{hughes2019largest}, the best optimizers are based on the Voronoi tessellation \cite{toussaint1983computing,chazelle1993optimal,okabe1997locational}, which is computationally prohibitive in high dimensions, requiring $O(n^{\lceil d/2 \rceil})$ operations for $n$ constraints.
The VA-edge class provides a geometrically intuitive and computationally efficient approach to this problem.
By analogy to simulated annealing, the reluctant VA-min algorithm can indeed find robust optimal configurations for a modest computational cost, $O(n d^3)$ (SI Appendix).

\bibliography{biblioPNAS.bib}

\pagebreak
\onecolumngrid

\setcounter{secnumdepth}{2}
\pagebreak
\onecolumngrid
\noindent{\centering\large\bfseries Supplementary Material\par}
\appendix
\section{VA-edge Algorithm}\label{app:VAA}
\begin{figure}[!ht]
\centering
    \includegraphics[width=0.59\columnwidth]{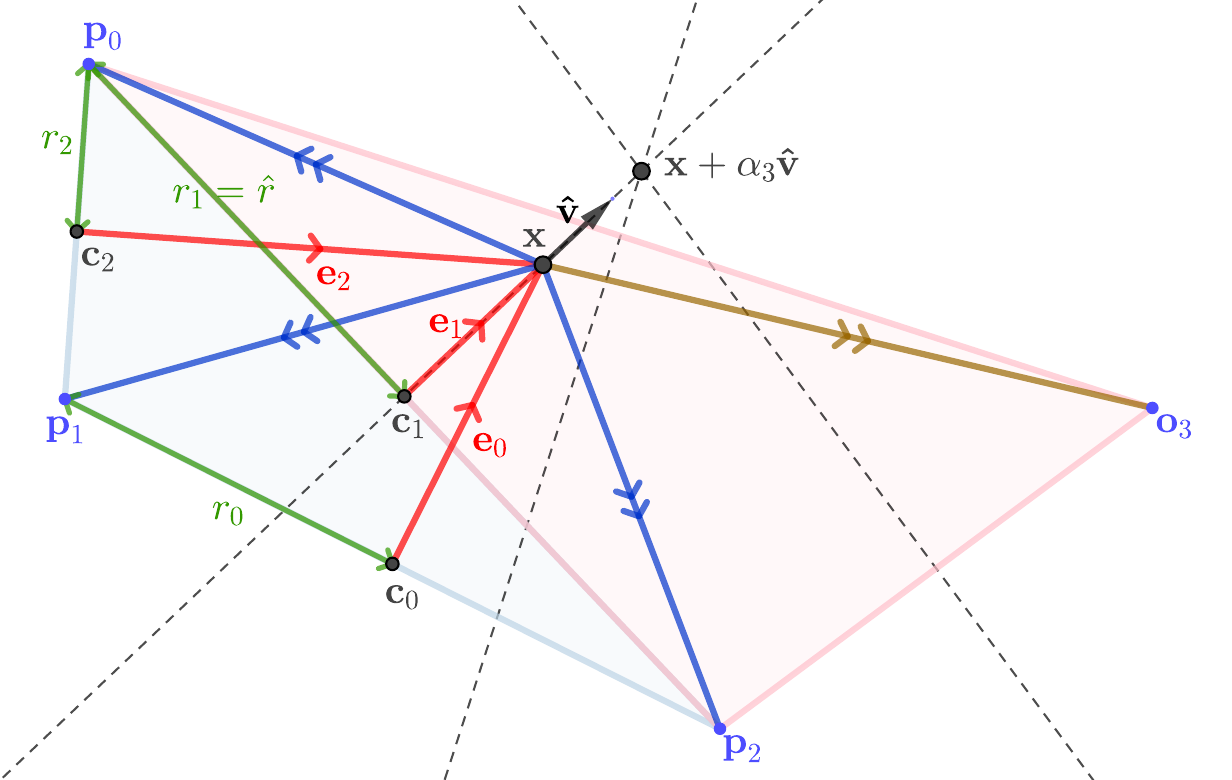}
    \caption{Geometry in $d=2$ of one VA-max step between VVs as reported in part (3) of Algorithm~\ref{alg:cap}.}
    \label{fig:step3}
\end{figure}

The VA-edge algorithm consists of three parts: (1) initializing the system by sampling $M$ points (corresponding to obstacle centers) uniformly at random within a spherical shell centered around the origin; (2) initiating the dynamics from the origin in the direction radial to the closest center, thus defining a Voronoi polytope, and proceeding for $d$ projection steps on the Voronoi faces until reaching one vertex of the polytope; (3) evolving the dynamics on edges of the Voronoi tessellation until reaching a stable vertex, i.e., an IS (see Fig.~\ref{fig:step3}). Details are provided in Algorithm~\ref{alg:cap}. The updated algorithm repository is available in \cite{VArep}.

\begin{algorithm}
\small
\caption{VA-edge algorithm}\label{alg:cap}
\begin{enumerate}
    \item Radially sample $M$ points $\{\mathbf{o}_i\}_{i=1}^{M}$ in $d$ dimensions around the origin $(0,\cdots,0)$.
    \item Project onto one edge of the initial Voronoi polytope.
    \begin{enumerate}
        \item Select the closest center $\mathbf{p}_0$, which is the center of initial Voronoi polytope.\\
                $\qquad \mathbf{p}_0=\text{argmin}_{\mathbf{o}_i}(|\mathbf{o}_i|)$
        \item Initiate the dynamics at the origin $\mathbf{x}_\mathrm{in}=\mathbf{x}_0=(0,\cdots,0)$ with an initial unitary velocity radial from $\mathbf{p}_0$.\\
                $\qquad \mathbf{\hat{v}}_0=-\frac{\mathbf{p}_0}{|\mathbf{p}_0|}$
        \item Proceed by projecting $\mathbf{v}_0$ on hyperplanes defined by successive closest points $\mathbf{p}_j$.\\
                \begin{algorithmic}
                    \For {$j=1; j < d+1; j+\!+$}
                        \For {$i=0; i < M; i+\!+$}
                            \If {$\mathbf{o}_i \notin \{\mathbf{p}_k\}_{k=0}^{j-1}$}\algorithmiccomment{if point does not already define a VV}
                                \State $l_i = \frac{|\mathbf{o}_i|-\mathbf{x}_{j-1}  \cdot  \mathbf{o}_i}{\mathbf{\hat{v}}_{j-1} \cdot \mathbf{o}_i}$\algorithmiccomment{distance of $\mathbf{x}_{j-1}$ in $\mathbf{\hat{v}}_{j-1}$ direction from new $\mathbf{o}_i$}
                            \EndIf
                        \EndFor
                    
                    \State $\mathbf{p}_j=\text{argmin}_{\mathbf{o}_i}(l_i)$ \algorithmiccomment{closest point in $\mathbf{\hat{v}}_{j-1}$ direction}
                    \State $\mathbf{x}_{j} = \mathbf{x}_{j-1} + l_i \mathbf{\hat{v}}_{j-1}$ \algorithmiccomment{update the position}
                    \State $\mathbf{v}_{j} = \mathbf{\hat{v}}_{j-1}-\mathrm{P}_{\{\mathbf{p}_k\}_{k=0}^{j}}\mathbf{\hat{v}}_{j-1}$ \algorithmiccomment{project velocity onto space $\perp$ to space spanned by $\{\mathbf{p}_k\}_{k=0}^{j}$}
                    \State $\mathbf{\hat{v}}_{j} = \text{sign}(\mathbf{v}_{j}\cdot \mathbf{x}_{j-1})\frac{\mathbf{v}_{j}}{|\mathbf{v}_j|}$ \algorithmiccomment{define the expanding orientation and renormalize the velocity}
                    \EndFor
                    \State $\mathbf{x}_\mathrm{VV} = \mathbf{x}_d = \mathcal{C}(\{\mathbf{p}_k\}_{k=0}^{d})$ \algorithmiccomment{the first VV is found, i.e. the circumcenter $\mathcal{C}$ equidistant to $\{\mathbf{p}_k\}_{k=0}^{j}$}
                \end{algorithmic}
    \end{enumerate}
    \item VA walk on Voronoi vertices until a stable one is identified.\\
        \begin{algorithmic}
            \While{$\mathcal{C}(\{\mathbf{p}_k\}_{k=0}^{d}) \notin \mathcal{S}(\{\mathbf{p}_k\}_{k=0}^{d})$}  \algorithmiccomment{check if stable, i.e. if $\mathcal{C}$ is contained in corresponding simplex $\mathcal{S}$}
            \State $\mathbf{x}=\mathcal{C}(\{\mathbf{p}_k\}_{k=0}^{d}), r = \mathcal{R}(\{\mathbf{p}_k\}_{k=0}^{d})$  \algorithmiccomment{define actual VV and respective radius (aka $r_\mathrm{sph}$)}
            \For {$j=0; j < d;j+\!+$}\algorithmiccomment{find expansion directions}
                \State $\mathbf{c}_{j}=\mathcal{C}(\{\mathbf{p}_k\}_{k\neq j}^{d}), r_{j}=\mathcal{R}(\{\mathbf{p}_k\}_{k\neq j}^{d})$\algorithmiccomment{define circumcenter and radius of $(d-1)$-simplex when excluding point $\mathbf{p}_j$}
                \State $\mathbf{e}_{j}=\mathbf{x}-\mathbf{c}_{j}$ \algorithmiccomment{define Voronoi edge vector when excluding point $\mathbf{p}_j$}

                \If{$\mathbf{e}_{j}\cdot (\mathbf{p}_j-\mathbf{x}) <\mathbf{e}_{j}\cdot  (\mathbf{p}_{*(\neq j)}-\mathbf{x})$}\algorithmiccomment{growth on edge $j$ from $\mathbf{p}_j$ must be smaller than for some $*\neq j$}
                    \State $\text{append}\quad  j \quad \text{to} \quad J$ \algorithmiccomment{$J$ is the list of expanding edges}
                \EndIf
            \EndFor
            \State $k_{\text{max}}=\text{argmax}_{k\in J}(\frac{|\mathbf{e}_{j}|}{r_{j}})$\algorithmiccomment{define direction of maximal expansion for VA-max (alternatively, argmin for VA-min)}
            \State $\mathbf{\hat{v}} = \frac{\mathbf{e}_{k_{\text{max}}}}{|\mathbf{e}_{k_{\text{max}}}|}, \hat{r} = r_{k_{\text{max}}}$\algorithmiccomment{unit vector along edge of maximal expansion and respective radius}
            \For {$i=0; i < M; i+\!+$}
                \If {$\mathbf{o}_i \notin \{\mathbf{p}_k\}_{k=0}^{d}$}\algorithmiccomment{if point does not already define VV}
                    \State $\alpha_i = \frac{|\mathbf{o}_i-\mathbf{x}|^2-\hat{r}^2}{2 \mathbf{\hat{v}} \cdot (\mathbf{o}_i-\mathbf{p}_{k\neq k_{\text{max}}}))}$\algorithmiccomment{$\alpha_i$ is such that $\mathbf{x} +\alpha_i \mathbf{\hat{v}}$ is equidistant from $\{\mathbf{p}_k\}_{k\neq k_{\text{max}}}^{d}$ and $\mathbf{o}_i$}
                \EndIf
            \EndFor
            \State $i_{\text{new}}=\text{argmin}_{i}(\alpha_i>0)$ \algorithmiccomment{select first point met expanding along the edge $\mathbf{\hat{v}}$}
            \State $\mathbf{p}_{k_{\text{max}}} = \mathbf{o}_{i_{\text{new}}}$ \algorithmiccomment{substitute new point thus defining the new VV}  
            \EndWhile
        \end{algorithmic}
        
\end{enumerate}
\end{algorithm}

\section{VA-edge Computational Efficiency}
\label{app:VAB}

\begin{figure}[!ht]
\centering
    \includegraphics[width=0.49\columnwidth]{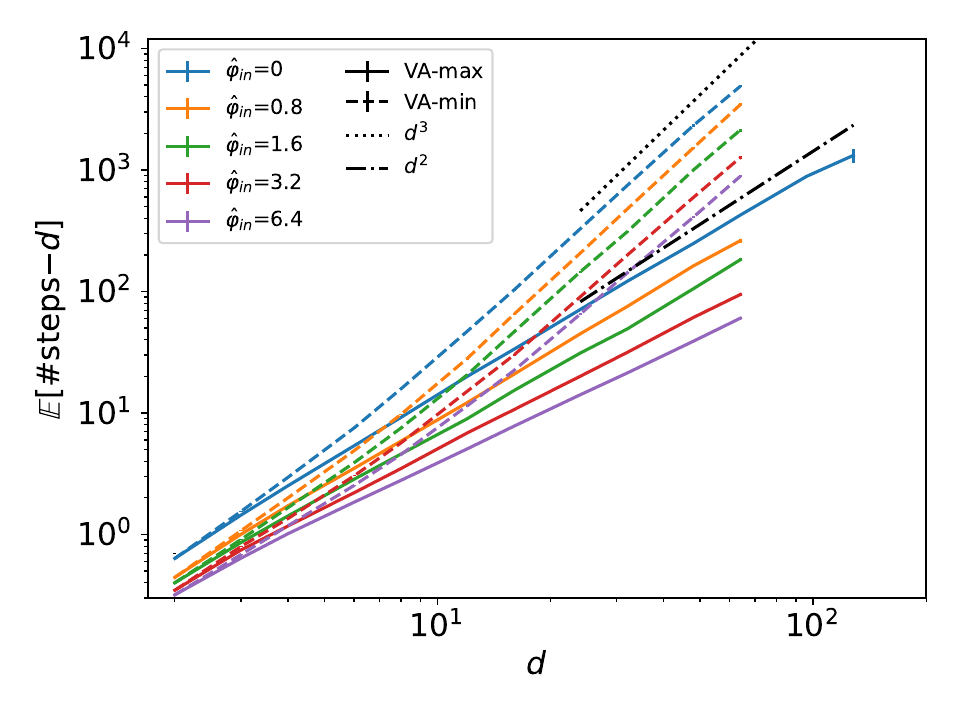}
    \includegraphics[width=0.49\columnwidth]{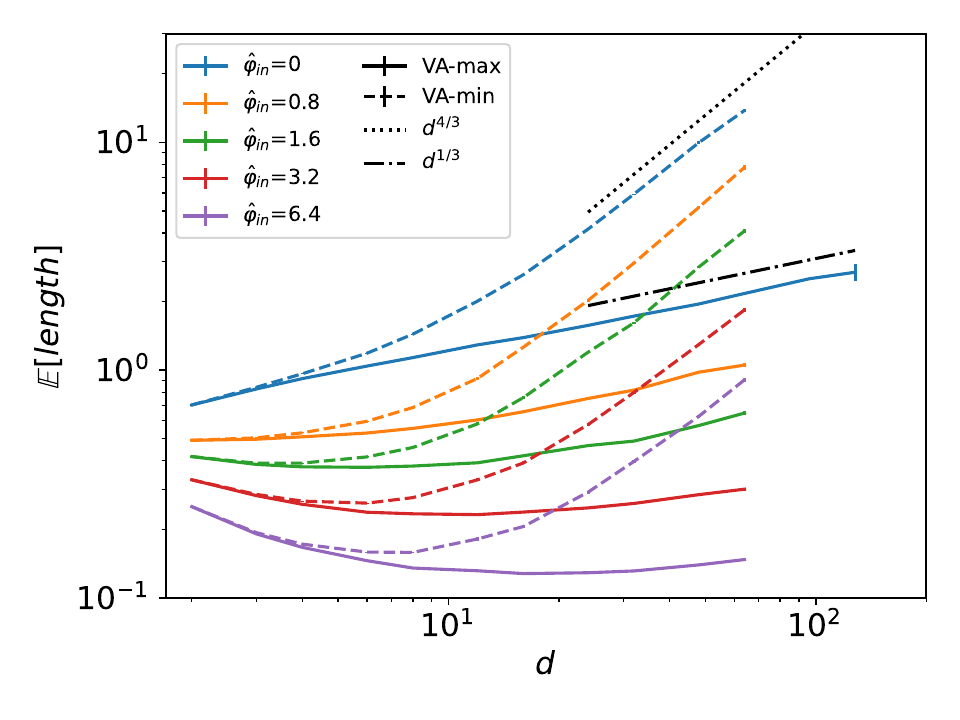}
    \caption{(left): Dimensional evolution of the average number of steps (after the $d$ steps of the initial projection) to reach an IS for $\hat{\varphi}_\mathrm{in}=0,0.2,0.4,0.8,1.6$. VA-max (full-line) reaches an IS in $O(d^2)$ steps, and VA-min (dashed-line) does so in $O(d^3)$ steps.
    (right): Average length of the total trajectory of the sph center to reach the IS. This quantity is proportional to the total time if the sph center evolves at a constant velocity.}
    \label{fig:RLGStepLen}
\end{figure}

This appendix presents the dimensional scaling of several observables from VA-max and VA-min for various $\hat{\varphi}_\mathrm{in}$. First, consider the total number of steps and the total displacement of the center of sph, i.e.\ the length of the trajectory. As can be seen in Fig.~\ref{fig:RLGStepLen}, both quantities asymptotically increase with dimension and decrease with $\hat{\varphi}_\mathrm{in}$. Moreover, the computational complexity of VA-max, $O(d^2)$, is always smaller than that of VA-min, $O(d^3)$. The low-polynomial scaling of VA-edge algorithms is non trivial. The number of unstable DS (or VVs) grows exponentially with $d$, and therefore a random walk on edges would take an exponential number of steps to reach an IS, as in typical NP-hard problems.  The VA constraint of increasing $r_\mathrm{sph}$, however, poses a strong ordering to the exploration, thus making VA-edge algorithms run in polynomial time in the dimension of the explored space.

The computational problem of finding an IS of large $\hat{\varphi}$ is a common optimization problem, very similar in spirit to optimization problems that involve minimizing a cost or energy function. Our problem is defined in a high-dimensional space when $d \gg 1$, and the structure of solutions, i.e.\ ISs, is strongly non-convex. This setup is typical of the situations in which algorithms can easily get trapped in suboptimal solutions and the optimal solution is hard to achieve (and sometimes even to approximate).

We have observed \comment{in Fig.~\ref{fig:Fig5}(a)}
that different algorithms reach ISs with different $\hat{\varphi}$. More importantly, we observe that for any $d$ VA-max and CALiPPSO reach seemingly equal densities 
as do VA-min and force-min. 
Two different algorithms reaching the same IS for any $d$  is hardly a coincidence and suggests that $\hat{\varphi}_\text{min}$ and $\hat{\varphi}_\text{max}$ play important roles. Could they be threshold values for different classes of algorithms?

To try to answer to the above question, we recap what is known about the paradigmatic model for complex systems, the Sherrington--Kirkpatrick (SK) model. The model is composed of $N$ Ising spins interacting through randomly chosen couplings (e.g.\ Gaussian couplings of zero mean and variance equal to $1/N$). In the large $N$ limit, the ground state energy is known from the Parisi solution, but it is not easily attainable by a polynomial-time algorithm. For example, it is well-known that greedy algorithms (i.e.\ algorithms decreasing the energy at each step, like the VA algorithms) reach different asymptotic energies depending on the degree of greediness. The most greedy version gets trapped far away from the ground state, while the most reluctant version more closely approaches the ground state~\cite{parisi2003statistical}. Recently, Montanari \cite{Montanari2019} presented an algorithm that approximates the ground state energy of the SK model to arbitrary precision in a time that scales quadratically with the system size (and inversely in the precision). Even more recently, the most reluctant algorithm has been reanalyzed \cite{erba2024}, finding evidence it can approach the ground state energy in a time $O(N^2)$ (while the most greedy version runs in a time $O(N)$). The overall picture that comes out from the study of algorithms optimizing the SK model is that different energy thresholds exist for algorithms running in times scaling differently with system size, that is, with the dimension of the space over which the function to be optimized is defined.

Our results on the computational complexity of the algorithms belonging to the VA-edge class fit perfectly into the above scenario. We have shown that the VA-max algorithm runs in a time $O(d^2)$ and reaches a lower value $\hat{\varphi}_\text{min}$. Instead, the VA-min algorithm runs in a time $O(d^3)$ and reaches a much larger value $\hat{\varphi}_\text{max}$.
Given the above observations on the universality of these two threshold values, we are tempted to conjecture that they correspond to algorithmic thresholds for classes of VA algorithms running on different time scales.
Moreover, given that the difference between these two thresholds gets larger as $d$ increases, the two algorithmic thresholds are well separated in the limit $d\rightarrow\infty$.

The linear optimization algorithm CALiPPSO converges in a number of steps $O(d)$. Notice, however, that each step is non-local and costs at least $O(d^2)$ operations. The overall complexity therefore scales as $O(d^3)$. Given that CALiPPSO and VA-max reach the same densities but that the latter takes a time $O(d^2)$, we conclude the class of VA-edge algorithms is highly efficient compared to other optimization algorithms.

\begin{figure}[!ht]
\centering
		\includegraphics[width=0.49\columnwidth]{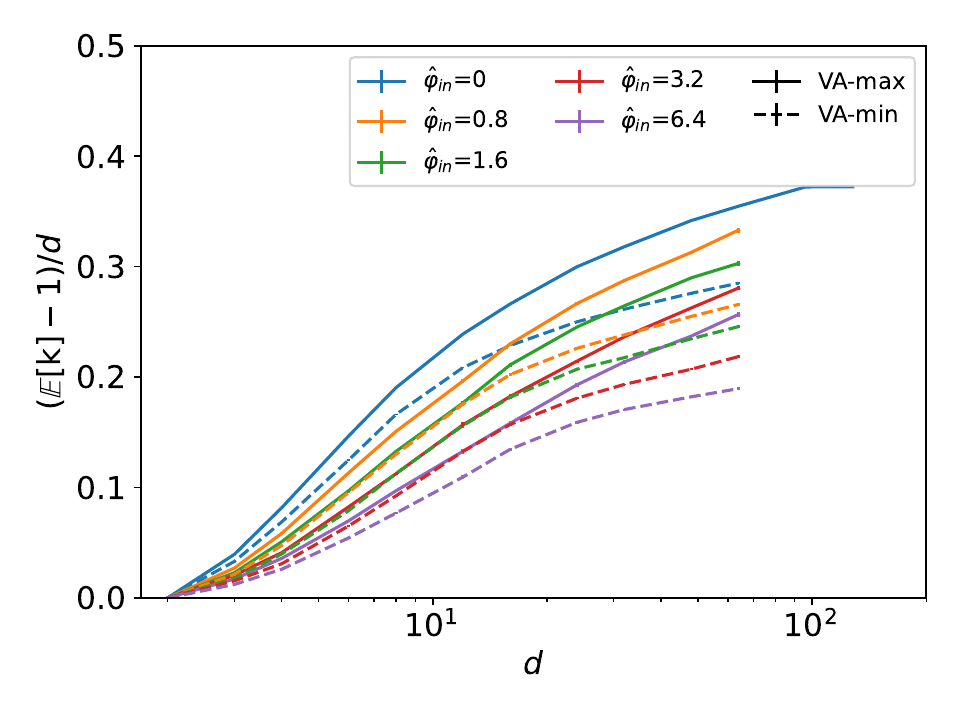}
        \includegraphics[width=0.49\columnwidth]{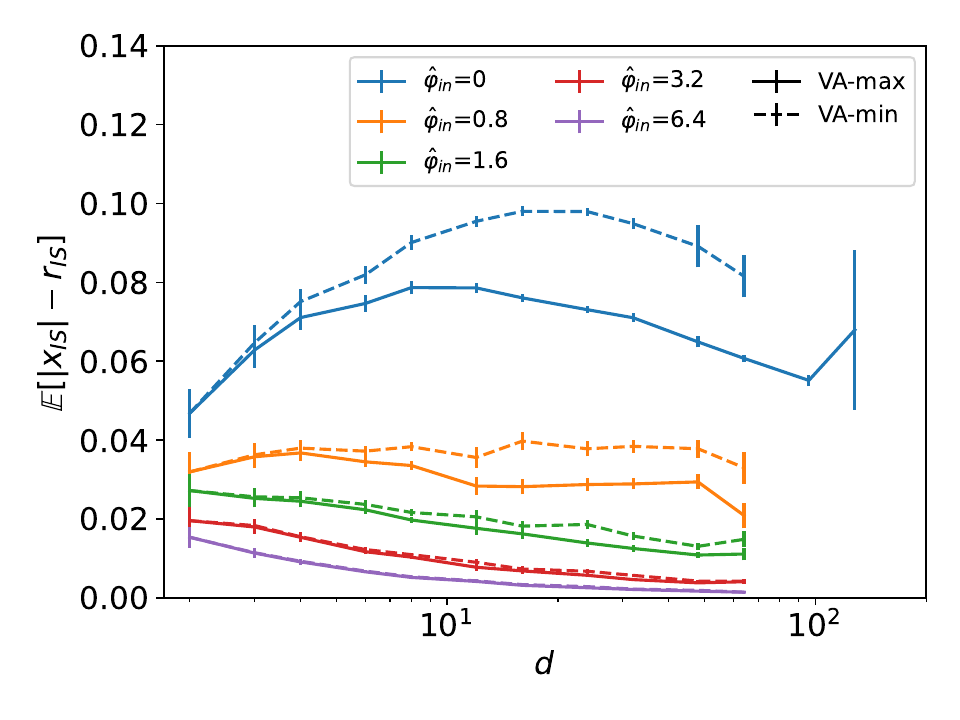}
		\caption{(left): Average proportion of expanding directions (minus one) over the VA trajectory as a function of $d$ for different $\hat{\varphi}_\mathrm{in}$.  VA-max (full-line) experiences more possible expansion directions than VA-min (dashed-line). 
        (right): Average difference between the final displacement of the sph center $|\mathbf{x}_\mathrm{IS}|$ and its final radius, $r_\mathrm{IS}$. That difference is zero if the final IS has a $r_\mathrm{IS}$ that touches the closest initial obstacle to the tracer. The behavior for different $\hat{\varphi}_\mathrm{in}$ suggests an asymptotic limit in which $r_\mathrm{IS}$ always touches the first encountered obstacle.}
  \label{fig:RLGEdgeMSDmR}
\end{figure}

During the VA-edge dynamics at each vertex the algorithm has $k$ different possible edges on which to expand. This number can be averaged during the whole trajectory for each trajectory. The proportion of expanding edges (i.e. $k/d$) is shown in Fig.~\ref{fig:RLGEdgeMSDmR} to asymptotically increase with dimension and decrease with $\hat{\varphi}_\mathrm{in}$. Note that the VA-max algorithm always sees more expanding directions, a signature of the fragility of trajectory and the final IS relative to those of VA-min. A similar result is expected for other reluctant VA-edge algorithms. 

The difference between the final displacement of the sph center $|\mathbf{x}_\mathrm{IS}|$ and its $r_\mathrm{IS}$ is shown in Fig.~\ref{fig:RLGEdgeMSDmR}. This quantity is identically zero if the final IS has a radius that is in contact with the first encountered obstacle. For asymptotically large $d$, this difference goes to zero for any $\hat{\varphi}_\mathrm{in}$. In other words, in the limit $d\to\infty$ the dynamics is confined to the Voronoi polytope in which it started. Studying the distribution properties of Poisson--Voronoi tessellations for large $d$ should therefore suffice to understand the fate of the VA-edge dynamics. Perhaps an analytical understanding of $\hat{\varphi}_\mathrm{J0}$ in the RLG through stochastic geometry 
would also be possible.

\section{CALLiPSO and the RLG}\label{app:CAL}
The definition of Delaunay basins 
\comment{in Eq.~\ref{eq:DelB}}
can be recast in a computationally powerful -- albeit non-local VA -- algorithm. In this scheme, the initial tracer position is first mapped to the circumcenter of the DS to which it belongs. That DS is either stable (i.e. an IS) or not. In the unstable case, the procedure is iterated until a stable DS is reached. (The computational complexity of the linear optimization algorithm arises not from evaluating the circumcenter, but from identifying which DS contains the point.) 
Each step is a linear optimization (LO) problem, and the resulting algorithm in jamming of multi-particle systems has been called CALiPPSO (for chain of approximate linear programming for packing spherical objects)~\cite{artiaco2022}.

For the RLG, each step of the CALiPPSO algorithm consists of finding the circumcenter corresponding to the vertices of the Delaunay simplex that contains the tracer (i.e. finding the displacement of the tracer $\mathbf{x}_\mathrm{circ}$ that most increases its radius $r_\mathrm{circ}$), then updating the tracer position to this new center. Therefore, each step corresponds to $\max_{\mathbf{x}_\mathrm{circ},r_\mathrm{circ}} r_\mathrm{circ}$, such that 
$r_\mathrm{circ} \leq |r_i-\mathbf{x}_\mathrm{circ}|^2  \quad \forall i$. This scheme can be translated into the LO problem:
\begin{equation}
\begin{cases}
\max_{\mathbf{x}_\mathrm{circ},\Gamma_\mathrm{circ}} \Gamma_\mathrm{circ}\quad \text{s.t.} \\
2\frac{\mathbf{r}_i}{|\mathbf{r}_i|}\cdot \mathbf{x}_\mathrm{circ} + \frac{\Gamma_\mathrm{circ}}{|\mathbf{r}_i|}\leq |\mathbf{r}_i|  \quad \forall i
\end{cases}
\end{equation}
where $\mathbf{r}_i$ is the position of the obstacle $i$ with respect to the tracer, $\mathbf{x}_\mathrm{circ}$ is the position of the circumcenter of the Delaunay simplex containing the tracer, and $\Gamma_\mathrm{circ}$ is a proxy for its radius $r_\mathrm{circ}=\sqrt{\Gamma_\mathrm{circ}+|\mathbf{x}_\mathrm{circ}|^2}$.
The computational complexity of each step is larger than $O(d^2)$. Note that this algorithm is equivalent to that described in \cite{artiaco2020,artiaco2022}, and here simply re-expressed for the RLG, in which case it becomes geometrically interpretable in terms of DS and corresponding circumcenters.

\section{Near-Equivalence of VA-max and CALiPPSO} \label{app:Equiv}
\begin{figure}[!ht]
        \centering
		\includegraphics[width=0.99\columnwidth]{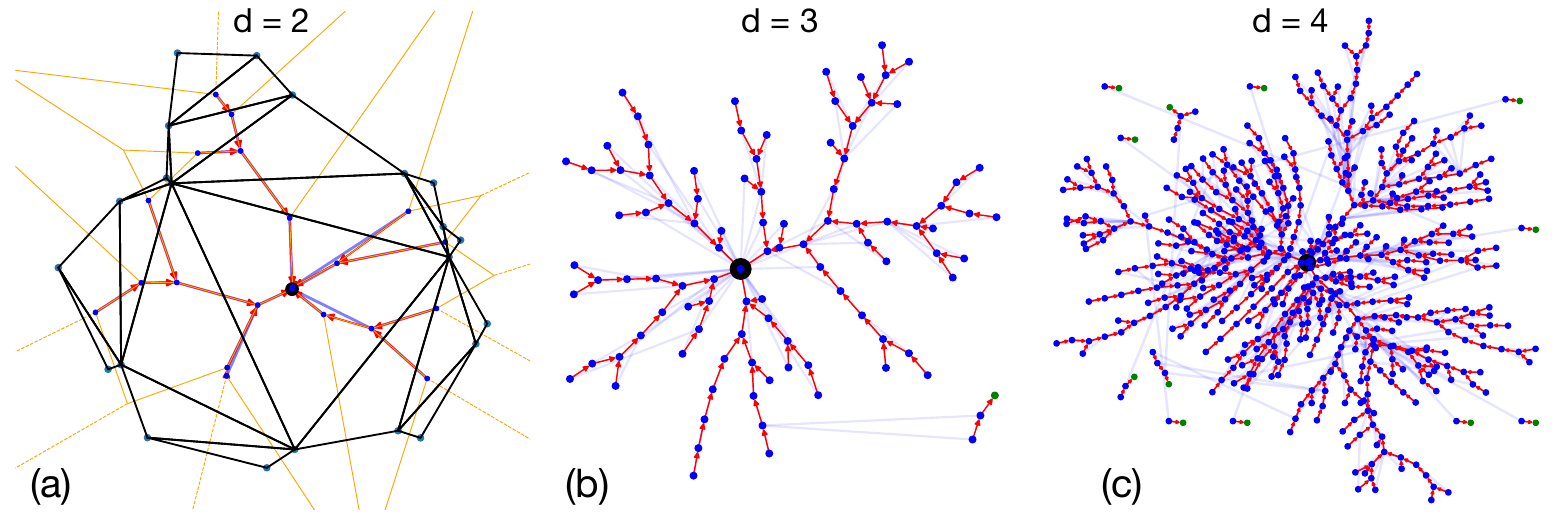}\\
		\caption{Comparison of CALiPPSO and VA-max within a large basin for $d=2$, 3, and 4. (a) For $d=2$, we display the graph embedded in real space along with the underlying Voronoi tessellation. Each blue dot marks the circumcenter of a simplex forming the Delaunay basin, connected by light blue lines as defined by the CALiPPSO algorithm. Red arrows indicate the VA-max trajectories. The black dot at the center represents the stable circumcenter, i.e., the IS. In $d=2$, the paths of both algorithms generally coincide, though CALiPPSO occasionally skips a Voronoi vertex, creating minor local discrepancies. These discrepancies become more pronounced in (b) $d=3$ and (c) $d=4$. The overall basin structure nevertheless remains largely consistent. Most blue dots follow VA-max (red arrows) and connect to other blue dots. A few exceptions (green dots) fall outside the Delaunay basin and may lead to different ISs.   }
  \label{fig:CALvsVAmax}
\end{figure}

VA-max and CALiPPSO are -- at least seemingly -- significantly different algorithms. Their associated basins of attraction in the RLG, however, are nearly identical. (i) VA-max follows an initial projection dynamics that has no equivalent in CALiPPSO. (ii) Both VA-max and CALiPPSO follow a VV-based dynamics, but VA-max follows the VA-edge that locally maximizes tracer growth, while CALiPPSO chooses the nearby VV that maximizes the growth of the tracer, hence directly optimizing the end point of each step. Both algorithms nevertheless follow a greedy path over VVs. For the RLG in $d=2$, if we exclude the initial projection of the VA-edge, the two schemes are identical. But even in higher $d=3,4$ the respective basins of attraction are fairly similar (see Fig.~\ref{fig:CALvsVAmax}).  We conjecture that the two algorithms become equivalent in the $d \to \infty$ limit. This is supported by the convergence of their jamming packing fractions at large $d$ (Fig. 5(A), main text) and the similarity of their force geometries already observed at low $d$ (Fig. 6(B)).
In any case, it is unclear if this correspondence holds also for multi-particle systems.

\section{force-min and the RLG}\label{app:for}
One can also consider the force-min algorithm, which is an adaptation of the overdamped event-driven algorithm first described in Ref.~\cite{lerner2013simulations}, wherein a hard particle system approaches jamming through athermal compression \cite{charbonneau2023jamming}. Adapting that algorithm to the RLG is conceptually and computationally straightforward; the result is part of the VA class. After the projection phase, the choice of direction within the VA cone is not based on the rate of volume increase along a VA-edge, but on the magnitude of the forces applied by the obstacles on the tracer. If a VV is unstable, then at least one of these forces is negative. The force-min algorithm prunes the contact with the largest negative force recursively until all forces are positive. If more than one force is removed, then an infinitesimal VA might cause the tracer to overlap with an obstacle. In that case, the removed contact that causes the largest overlap upon infinitesimal ascent is reconnected, and the algorithm returns to pruning forces. This procedure is followed until an IS is identified.

If no more than one VV is removed, then the scheme is part of the VA-edge class. Otherwise, the trajectory goes through a Voronoi facet of larger dimension and is hence not part of that class. The final packing fraction nevertheless falls near the VA-min results\comment{(see Fig.~\ref{fig:Fig5}(a))}. A schematic recapitulating the various algorithms considered in this work is provided in Fig.~\ref{fig:SCH}.

\section{Overview of the Algorithms Used} \label{app:Class}
\begin{figure}[!ht]
        \centering
		\includegraphics[width=0.50\columnwidth]{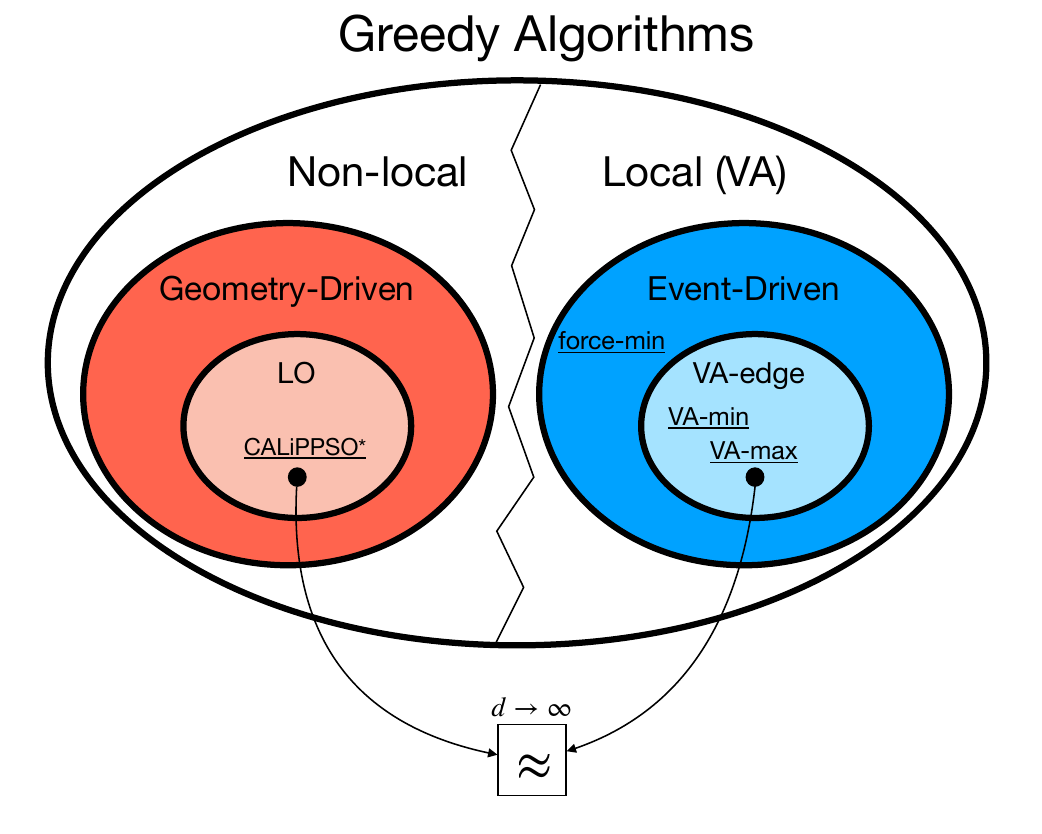}\\
		\caption{Scheme of the different classes of greedy algorithms for the RLG explored in this manuscript.}
  \label{fig:SCH}
\end{figure}

Figure~\ref{fig:SCH} summarizes the different classes of greedy algorithms for the RLG explored in this article. On the left, in the class of geometry-driven algorithms, the linear-optimization (LO) subclass approximates (at each step) the local geometry of growth with a convex polytope and update the position by linear optimization. For the RLG, the greediest known algorithm for this class is the CALiPPSO algorithm \cite{artiaco2020,artiaco2022}. On the right, in the class of event-driven local volume-ascent (VA) algorithms, the VA-edge subclass follows Voronoi edges to search for growing directions. For the RLG, by construction the greediest algorithm for this class is the VA-max algorithm, which is analogous to a gradient descent algorithm for energy landscapes. The force-min algorithm developed in \cite{lerner2013low} reaches denser packings than the VA-max and is therefore a sub-greedier (more reluctant) version. From the numerical simulations, we conjecture that VA-max and CALiPPSO are very similar in small $d$ and asymptotically equivalent in large $d$ (see discussions in Appendix \ref{app:Equiv}).

\section{Numerical Methods for the Geometrical Analysis}\label{app:numerical}

Each numerical simulation in this manuscript relies on sampling $M$ Poisson-distributed points (obstacles) with $\rho=1$ inside a $d$-dimensional ball of radius $R$. This construction is achieved by radial sampling, following the approach described in Ref.~\cite{Bonnet2024}. The choice of radius, $R = (M/V_d)^{1/d}$, is inherently related to the number of points $M$ and the dimension $d$. In general, $M$ is chosen based on the type of simulation performed.

\begin{figure}[!ht]
        \centering
		\includegraphics[width=0.74\columnwidth]{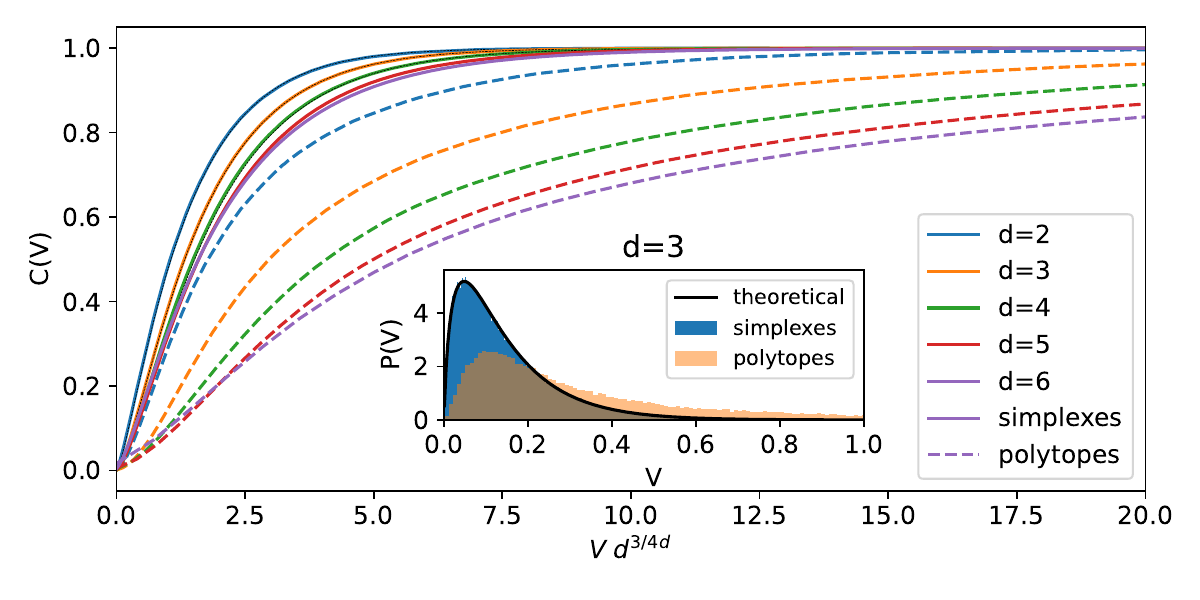}
		\caption{Cumulative distribution of the volume of simplexes (continuous) and polytopes (dashed) in the Poisson--Delaunay tessellation for $d=2,\dots,6$. The abscissa is rescaled for compactness. The polytopes (Delaunay basins) identify the basins of IS for $\varphi_\mathrm{in}=0$ obtained by agglomerating the stable simplex with the surrounding unstable ones. For simplexes, the exact mathematical result is superimposed (thin dotted black line); no comparable result exists for these polytopes. The inset shows the corresponding probability distributions for $d=3$.}
  \label{fig:RLGVolume}
\end{figure}

The geometrical properties of IS and corresponding Delaunay basins in the RLG \comment{-- shown in Figs.~\ref{fig:Fig2} and ~\ref{fig:Fig4} and in the red dotted lines in Fig.~\ref{fig:Fig2}(b) --} are evaluated numerically by exact Voronoi/Delaunay tessellation of $50$ samples (using qhull~\cite{qhull} as in \cite{Bonnet2024}). These samples contain $M = 2000 \cdot d$ points for $d=2,\dots,5$, and $M = 40,000$ points for $d=6$. Due to the exponential scaling of the computational complexity with dimension, simulations are limited to $d \leq 6$. The results align well with the analytical predictions of the radius and volume distribution of simplexes, as discussed in Appendix \ref{app:fractal} (see \comment{Fig.~\ref{fig:Fig2} and} Fig.~\ref{fig:RLGVolume}).

For the compression analysis of different algorithms, the radius $R$ is selected so that $M = 1000d^2$ points are sampled. This scaling has been empirically found necessary to reach dimensions as high as $d = 200$. Given that the reduced packing fraction $\hat{\varphi}$ is expected to converge to a finite value, and since $d\hat{\varphi}_\mathrm{IS}$ represents the volume of a sphere with radius $r_\mathrm{IS}$, and $M$ equals the total volume of the simulation ball (since $\rho=1$), we expect to need at least $M = O(d^1)$ for the simulation ball to contain the final sph. However, accounting for the displacement of the center of the sph, we have chosen $M = O(d^2)$.


\section{Delaunay Basin Volume and Geometry}\label{app:fractal}

The Poisson--Voronoi/Delaunay tessellation offers a minimal model of random tessellations and has hence been intensively studied by mathematicians and computer scientists. For example, the volume distribution of Poisson--Delaunay simplexes~\cite{rathie1992volume} (see Fig.~\ref{fig:RLGVolume}) and the distribution of circumradii of Delaunay simplexes~\cite{edelsbrunner2017expected} are known exactly. These results have here been used to validate the RLG IS numerical simulations.
However, to the best of our knowledge, no studies of polytopes formed by the union of one stable simplex and the surrounding unstable ones have previously been made. Therefore, in order to study the results of the compression algorithm, numerical simulations remain the best option. In Fig.~\ref{fig:RLGVolume} we compare the distribution of the volume of Delaunay simplexes with the volume of Delaunay basins (polytopes).

As shown in Fig.~4 of the main text, the distribution of volumes versus IS radius of Delaunay basins follows an anomalous power law, reflecting their fractal nature. To quantify this, we estimate the correlation dimension \cite{theiler1990estimating}. For each Delaunay basin, we compute the radial cumulative volume distribution:

\begin{equation}
\mathcal{C}(r) = \frac{\sum_{s \; \text{s.t.}\; r_s<r } v_s}{\sum_{s} v_s} \ ,
\end{equation}
where $ v_s $ is the volume of simplex $ s $, and $ r_s $ is the distance of its barycenter from the barycenter of the entire basin. The sum $ \sum_{s} $ runs over all simplexes comprising the Delaunay basin. This formulation is an approximation as it assumes the volume of each simplex is concentrated at its barycenter, disregarding its actual shape. However, the validity of this approximation improves with increasing dimension.

\begin{figure}[!ht]
        \centering
		\includegraphics[width=0.99\columnwidth]{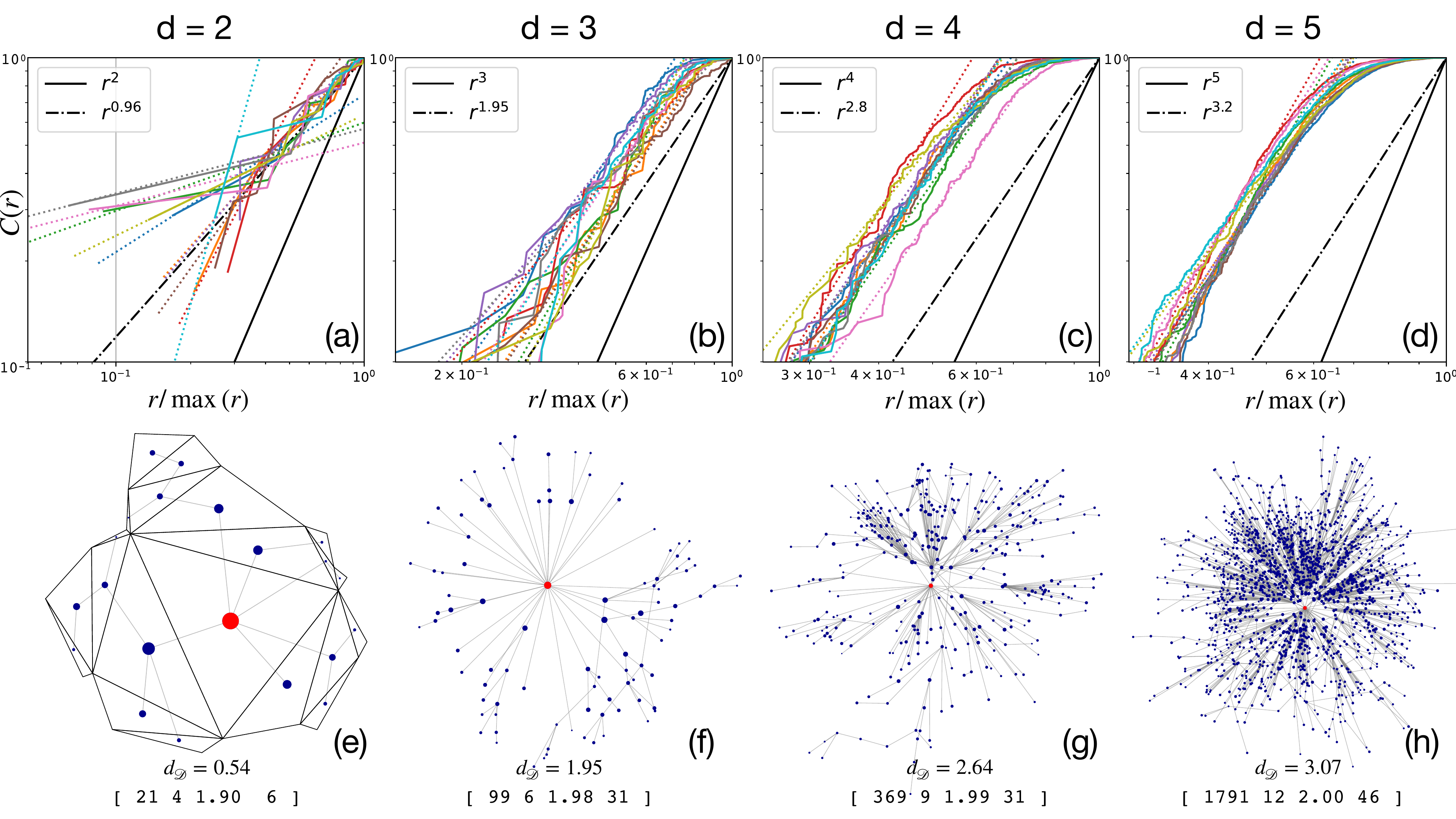}
		\caption{For dimensions $ d=2,\dots,5 $, the upper panels (a) to (d) show the radial cumulative volume distribution (colored lines) for the ten largest Delaunay basins in a simulated ball containing $ 20000d $ obstacles. The solid black line represents the expected growth for compact volumes, $ r^d $, while the dash-dotted line indicates the average fractal growth, $ r^{d_\mathcal{D}} $, with $ d_\mathcal{D} $ reported in Table~\ref{tab:fractal_dimension}. 
        Panels (e) to (h) illustrate the graph structure of the largest Delaunay basin for $ d=2,\dots,5 $. Each dot's area is proportional to the volume of its corresponding simplex. The red dot marks the central stable simplex. Below each graph, the corresponding fitted fractal dimension is provided, along with four graph metrics in square brackets [\#nodes, max depth, branching factor, max branching factor], which can be compared with typical values in Table~\ref{tab:tree_metrics}.}

  \label{fig:fract}
\end{figure}

We fit each distribution with a power law in the intermediate regime, specifically considering radii where the relative volume falls between 20\% and 60\% of the total volume. The results are presented in Fig.~\ref{fig:fract}. Panels (a) to (d) (for dimensions $ d=2,\dots,5 $) display the radial cumulative volume distribution (colored lines) for the ten largest Delaunay basins in a simulation with $ 20,000d $ obstacles. The corresponding power-law fits are shown as dotted lines. For compact volumes, the expected scaling is $ r^d $ (solid black line). However, we find that the basins exhibit a fractal dimension $ d_\mathcal{D} $, which is strictly smaller than $ d $. The average fractal dimension across the ten fits (dash-dotted line) is reported in Table \ref{tab:fractal_dimension}.
Note, however, that these fits are intended to qualitatively illustrate the emergence of long-range correlations in large basins, rather than to extract precise exponents. Alternative functional forms, such as exponential or stretched exponential decays, may offer complementary insights, but a systematic comparison is left for future work.

Figures (e) to (h) illustrate the tree graph structure of the largest basin among them, represented as in the main text. 
Table~\ref{tab:tree_metrics} quantifies the internal structure of Delaunay basin graphs across dimensions $d = 2$ to $5$. As $d$ increases, the graphs grow both in size and complexity. The average number of nodes, the average maximum depth and the average branching factor ---both mean and maximal--- all rise. These trends confirm that Delaunay basins become increasingly hierarchical and ramified in higher dimensions, consistent with the emergence of the fractal-like organization suggested by the radial volume analysis.

\begin{table}[ht]
    \begin{minipage}{0.49\textwidth}
        \centering
        \begin{tabular}{|c|c|c|c|c|}
            \hline
            $d$ & 2 & 3 & 4 & 5 \\
            \hline
            $d_\mathcal{D}$ & 0.96(5) & 1.95(8) & 2.8(2) & 3.2(3) \\
            \hline
        \end{tabular}
        \caption{Delaunay fractal dimensions}
        \label{tab:fractal_dimension}
    \end{minipage}
    \hfill
    \begin{minipage}{0.49\textwidth}
        \centering
        \begin{tabular}{|c|c|c|c|c|}
            \hline
            $d$ & 2 & 3 & 4 & 5 \\
            \hline
            $\langle$\# nodes$\rangle$ & 3.4 & 9.4 & 25.3 & 73.8 \\
            $\langle$max depth$\rangle$ & 1.4 & 2.3 & 3.2 & 4.3 \\
            $\langle$branching factor$\rangle$ & 1.0 & 1.4 & 1.6 & 1.8 \\
            $\langle$max branching factor$\rangle$ & 1.7 & 3.8 & 7.2 & 13.6 \\
            \hline
        \end{tabular}
    \caption{Delaunay graph metrics}
    \label{tab:tree_metrics}
    \end{minipage}
    \caption*{\textbf{(TABLE S1):} Fractal dimension of Delaunay basins estimated from fits of the radial distribution of volume shown in Fig.~\ref{fig:fract} for the first 10 largest basins for each $d$. \textbf{(TABLE S2):} Typical Delaunay graph structure metrics for $d=2,\dots,5$. Each property is averaged over Delaunay basins with a weight proportional to the volume of basin, i.e. $\langle O \rangle=\frac{\sum_{b} O_b v_b}{\sum_{b} v_b}$, where $v_b$ is the volume of the basin $b$.}
\end{table}

\section{Force Normalization}\label{app:forces}
\begin{figure}[!ht]\label{fig:normF}
        \centering
		\includegraphics[width=0.99\columnwidth]{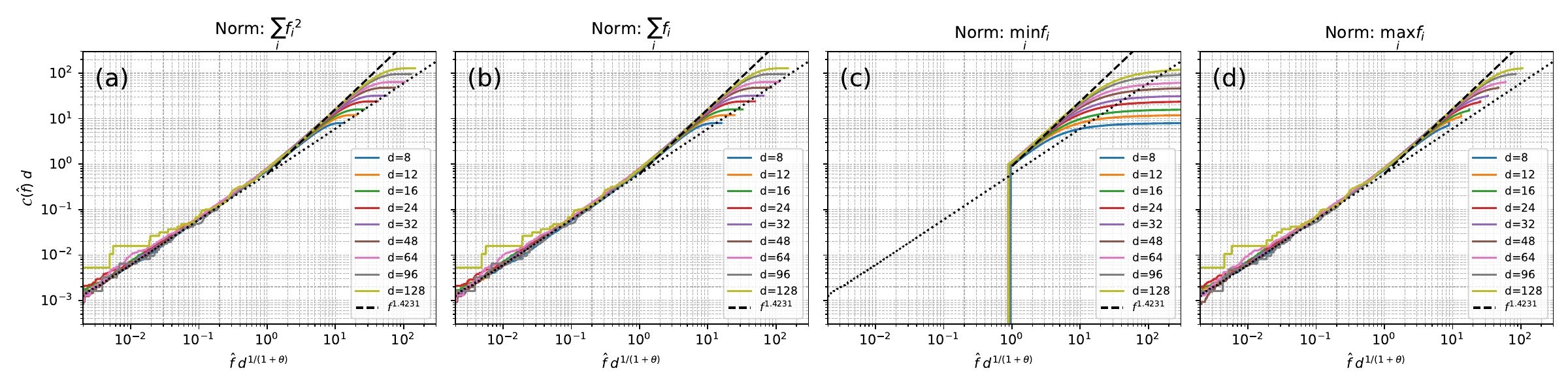}
		\caption{Cumulative distribution of the forces with different normalization for $d=8, 12, 16, 24, 32, 48, 64, 96, 128$: (a) normalization Eq.~\ref{eq:forcenorm} adopted in the main text; (b) linear version $\sum_{i=0}^{d} f_i = 1$; (c) $\min_i f_i =1$ and (d) is with $\max_i f_i =1$. All normalization are compatible with the power-law exponent in the limit $d\to\infty$, $1+\theta=1.4231...$.}
        \label{normF}
\end{figure}

The force distribution in the RLG model is extracted from the contact vectors $\mathbf{n}_i$. Each IS has $d+1$ contact vectors, with each vector connecting the position $\mathbf{x}_\mathrm{IS}$ to that of a surrounding obstacle, denoted $p_i$. For the IS to be mechanically stable, the forces exerted by these obstacles must balance out, leading to the condition \begin{equation}\label{eq:forcebalance} 
\sum_{i=0}^{d} f_i \mathbf{n}_i = 0 
\end{equation} 
where $f_0, \dots, f_d$ are the force magnitudes. This expression represents $d$ constraints for $d+1$ unknowns, implying one free scaling degree of freedom. To fix this freedom, we impose a normalization condition on the forces, 
\begin{equation}\label{eq:forcenorm}
\sum_{i=0}^{d} f_i^2 = 1 \ .
\end{equation}
Note that in multi-particle systems, such a particle-scale normalization is not required, because forces are then distributed and shared across the system.
In the main text, the lower tail of the force distribution is analyzed by constructing the joint histogram of all force magnitudes obtained from Eqs.~\ref{eq:forcebalance} and \ref{eq:forcenorm}.
Different normalization of the forces, e.g., using a power different than $2$ in Eq.~\ref{eq:forcenorm} or a reweighing of each force with the packing fraction reached at that IS are all compatible with the power-law exponent in the limit $d\to\infty$, $1+\theta=1.4231...$. Figure~\ref{normF} compares some of those normalizations.

\section{Finite-d Analysis of the Jamming Critical Distributions}\label{app:fss}

The analysis we carried out is completely analogous to the one presented in \cite{charbonneau2021finitesize} for the case of multi-particle jamming, so we here only describe it briefly. The main idea is that if in the thermodynamic limit a random variable $x$ follows a power-law pdf, $p_X(x) \sim x^\alpha$, for a given $\alpha$, sampling $x$ in finite systems will yield a distribution with a different exponent, say $\alpha'$. When $p_X$ is related to criticality, then the main reason for having $\alpha \neq \alpha'$ is not limited sampling but finite-size effects. That is, in finite systems with $n$ degrees of freedom, there is a natural bound to any correlation length in the system, which cannot be larger than the system itself, i.e. $\xi \sim L \sim n^{1/d}$, and therefore divergences in $\xi$ are inevitably suppressed.
This property, which initially seems a strong limitation, can actually be exploited to derive a rescaled variable $\tilde{x}$, which accounts for the effect of $n$ and $\alpha$ in the sampling of $x$, and the corresponding scaling function, $\tilde{p}_X(\tilde{x})$, which becomes independent of system size. Importantly, this is only the case when the correct value of $\alpha$ is used for defining $\tilde{x}$. In practical terms, only if we accurately estimate $\alpha$ do the empirical distributions of $p_X$ -- obtained from datasets of systems with different sizes -- collapse into a single master curve, i.e., the scaling function. 

Let us now consider the case of contact forces in the RLG, so $x=f$ and we know the $d\to \infty$ prediction is $p_F(f)\sim f^\theta$, with $\theta=0.4231\dots$. As explained in Ref.~\cite{charbonneau2021finitesize}, to derive the scaling function of interest we need to consider that the smallest force in a sample, $f_\mathrm{min}$ typically behaves as $\int_0^{f_\mathrm{min}} p_F(f) \text{d}f \sim f^{1+\theta}_\mathrm{min} \sim 1/n$. The next step is  realizing that,  while in multi-particle systems $n=dN$, in the RLG we have instead $n=d+1$. In other words, the correct variable for which to consider finite-size effects is dimensionality itself. 
With this mapping in mind, we can directly use the equations from Ref.~\cite{charbonneau2021finitesize}  and obtain the scaling variable for the forces, $\tilde{f} = f d^{1/(1+\theta)}$ and $\tilde{p}_F(\tilde{f})\sim d^{\frac{\theta}{1+\theta}} p_F(f)$. Additionally, size corrections to isostaticity are important because they make the pdf to $\tilde{p}_F(\tilde{f})\sim 1$ for $\tilde{f}\ll 1$. To better compare with numerical data in the main text we considered the cumulative distributions, $c(f)= \int_0^f p_F(f')\text{d}f'$,  for which the corresponding scaling function reads
\begin{equation}\label{eq:scaling-function-forces}
    \tilde{c}(\tilde{f}) \sim d c(f) \sim 
    \begin{cases}
    \tilde{f}^{1+\theta}, & \tilde{f} \gg 1\\
    \tilde{f}, & \tilde{f} \ll 1 \, .
    \end{cases}
\end{equation}
The first power-law scaling comes from the usual scaling regime in which finite-size effects are incorporated into the thermodynamic-limit criticality, and therefore explicitly depend on $\theta$; the linear scaling stems from finite $d$ corrections to isostaticity. Given that the results in Fig.~6(A) of the main text 
accurately follow both regimes of Eq.~\ref{eq:scaling-function-forces}, we conclude that jamming universality is clearly present in the RLG.

\end{document}